\documentclass[a4paper,11pt]{article}
\pdfoutput=1 %
\usepackage{jcappub} 

\usepackage[T1]{fontenc} 

\usepackage[utf8]{inputenc}

\usepackage{hyperref} 
\usepackage{url}
\usepackage{graphicx}
\usepackage{esvect}
\usepackage{graphics}
\usepackage{natbib}
\usepackage{mathrsfs}
\usepackage{blindtext}
\usepackage{comment} 
\usepackage{placeins} 
\usepackage{comment}

\usepackage[table]{xcolor}
\usepackage{multirow}

\usepackage[normalem]{ulem}
\usepackage{color}

\usepackage{soul}

\usepackage{supertabular}
\usepackage{tabularx}

\title{
The Cosmic-Ray sea explains the diffuse Galactic gamma-ray and neutrino emissions from GeV to PeV
\\
}

\author[a, b]{Pedro~De~La~Torre~Luque}
\emailAdd{pedro.delatorre@uam.es}
\author[c, d]{Daniele~Gaggero}
\emailAdd{daniele.gaggero@pi.infn.it}
\author[c]{Dario~Grasso}
\emailAdd{dario.grasso@pi.infn.it}
\author[e, f, g]{Antonio~Marinelli}
\emailAdd{antonio.marinelli@na.infn.it}
\author[h]{Manuel~Rocamora}
\emailAdd{manuel.rocamora@uibk.ac.at}

\affiliation[a ]{Instituto de Física Teórica, IFT UAM-CSIC, Departamento de Física Teórica,\\
Universidad Autónoma de Madrid, ES-28049 Madrid, Spain}
\affiliation[b ]{The Oskar Klein Centre, Department of Physics, Stockholm University, AlbaNova\\
  SE-10691 Stockholm, Sweden}
\affiliation[c]{INFN Sezione di Pisa, Polo Fibonacci, Largo B. Pontecorvo 3, 56127 Pisa, Italy}
\affiliation[d]{Dipartimento di Fisica, Universit\`a di Pisa, Polo Fibonacci, Largo B. Pontecorvo 3}
\affiliation[e]{Dipartimento di Fisica ``Ettore Pancini'', Università degli studi di Napoli ``Federico II'', Complesso Univ. Monte S. Angelo, I-80126 Napoli, Italy}
\affiliation[f] {INFN Sezione di Napoli, Complesso Univ. Monte S. Angelo, I-80126 Napoli, Italy}
\affiliation[g]{INAF- Osservatorio Astronomico di Capodimonte, Salita Moiariello 16, I-80131, Napoli, Italy}
\affiliation[h]{Universität Innsbruck, Institut für Astro- und Teilchenphysik, Technikerstr. 25/8, 6020 Innsbruck, Austria}

\date{\today}

\abstract{The LHAASO collaboration has recently released the spectrum and the angular distribution of the $\gamma$-ray Galactic diffuse emission from 1 TeV to 1 PeV measured with the Kilometer-2 Array (KM2A) and Water Cherenkov Detector Array (WCDA). 
We show that these data are in remarkably good agreement with a set of models that assume the emission to be produced by the Galactic population of cosmic rays if its spectral shape traces that measured by CALET and DAMPE as well as KASCADE at higher energies.
No extra-components besides the CR sea is needed to explain LHAASO results.
Accounting for unresolved sources, we consistently reproduce a wide set of $\gamma$-ray data at lower energy.
To do this, we consider two different transport setups: a conventional one and a $\gamma$-optimized spatial-dependent one (a development of the widely adopted KRA$_\gamma$ model). 
We demonstrate that both setups are compatible with LHAASO results. However, the latter is preferred if one takes into account Fermi-LAT gamma-ray data and neutrino measurements. 
In fact, we also compute the associated Galactic neutrino diffuse emission finding that the contribution from sources cannot be dominant and showing that spatial-dependent propagation models closely match the ANTARES and IceCube best fits for the Galactic Center Ridge and the Galactic Plane emissions. 
We argue that our $\gamma$-optimized model should be used as a template for future analyses of upcoming data from the Global Neutrino Network. 
}

\begin{document}
\maketitle
\flushbottom

\section{Introduction}
\label{sec:intro}



The Galaxy is a guaranteed source of neutrinos produced by the interaction of the dominant hadronic component of Galactic Cosmic Rays (CRs) with the interstellar medium gas.
This emission is the counterpart of the hadronically generated component of the $\gamma$-ray diffuse --  or more precisely, {\it interstellar} -- emission, detected by the Fermi Large Area space Telescope (LAT) \cite{Fermi-LAT:2012edv} from a fraction of GeV up to hundreds GeVs and, more recently, by several Air Shower, water Cherenkov and Cherenkov Array ground-based experiments at higher energies (see below).
The first pioneering estimates of the correlated expected neutrino Galactic diffuse emission ($\nu$GDE)~\cite{Berezinsky:1975zz,Stecker:1978ah} assumed uniform interstellar gas and CR and densities all over the Galaxy equal to their locally measured values.  More recent works considered more realistic gas~\cite{Berezinsky:1992wr} and, subsequently, CR source distributions determined on the basis of astronomical observations and computed the CR propagated spectrum in each point of the Galaxy solving the transport equation~\cite{Evoli:2007iy,Ahlers:2015moa}. Although predicting a significantly larger flux than the previous computations 
also such an improved approach generally found, however, a $\nu$GDE well below the expected sensitivities of the planned neutrino telescopes. 

Besides the improvements of experimental and analysis techniques, the search of the $\nu$GDE revived however by the introduction of a new class of CR transport models featuring a spatial-dependent CR transport setup suggested by a variety of analyses of Fermi-LAT data \cite{Gaggero:2014xla,Gaggero:2015xza,Fermi-LAT:2016zaq,Yang2016prd,Lipari:2018gzn,Pothast:2018bvh} and possibly connected to non-trivial aspects of CR transport microphysics  \cite{Cerri2017jcap,Recchia:2016bnd}. Such class of models typically predict a considerable increase of the $\nu$GDE in the innermost region of the Galactic Plane (GP) with respect to conventional (spatial-independent) models. 
In particular the so called KRA$_\gamma$ models \cite{Gaggero:2014xla,Gaggero:2015xza} were used in a number of ANTARES and IceCube analysis works looking for a possible signal from the GP~\cite{ANTARES:2017nlh,IceCube:2017trr}. Noticeably, a combined maximum-likelihood estimate of the data of both experiments \cite{ANTARES:2018nyb} yielded a non-zero $\nu$GDE flux for both models with a p-value of $29\%$ for KRA$_\gamma^5$ and $26\%$ for the KRA$_\gamma^{50}$. 
Afterwards the IceCube collaboration reported a $2\sigma$ hint using the KRA$_\gamma^5$ as a template with a best-fit normalization flux 0.85 times the predicted one~\cite{IceCube:2019lzm}.

The breakthrough, however, arrived with the IceCube collaboration announcement of a clear identification of the $\nu$GDE with a 4.5$\sigma$ significance~\cite{IceCube:2023ame}.
In order to cope with the overwhelming down-going muon background, a deep learning technique was used to build the sample of fifty thousand  neutrino cascade events used in this template fitting analysis. That required a statistical fit of the data against some spatial and spectral distribution template models of the emission. Again, the KRA$_\gamma^{5(50)}$ models (the exponent representing the assumed exponential cutoff energy of the CR source spectrum) were used for that purpose together with a conventional {\tt GALPROP}-based \cite{Strong:1998pw} model ( $ ^S {\rm S}^Z 4^R 20^T 150^C 5$  \cite{Fermi-LAT:2012edv}), identified as $\pi_0$ model in this context. That model does not include the aforementioned spatial-dependent diffusion setup.
Using that approach, IceCube \cite{IceCube:2023ame} found an excess with respect to the background with significances 4.71, 4.37 and 3.96 $\sigma$ for the $\pi_0$, KRA$_\gamma^{5}$ and KRA$_\gamma^{50}$ models, respectively. 
Even more recently, the KRA$_\gamma^{5}$ template was adopted as a benchmark in the analysis of the ultra-high-energy event KM3-230213A \cite{KM3NeT:2025npi} reported by the KM3NeT collaboration aimed at assessing its potential Galactic origin \cite{KM3NeT:2025aps}.

Regarding the IceCube analysis, it is important to notice the significantly different rescaling factors required for the original physical models to fit IceCube data. At 100 TeV they were found to be 4.5, 0.55 and 0.37 for the $\pi_0$, KRA$_\gamma^{5}$ and KRA$_\gamma^{50}$ models, respectively.
The large discrepancy between the $\pi_0$ IceCube fit and the original {\tt GALPROP}-based physical model may indicate a potential tension with Fermi-LAT data. 

%
{Moreover, the IceCube results show mild hints for a hardening of the $\nu$GDE in the innermost region of the GP, with $\vert l \vert < 30^\circ$, where the spectral index gets close to $- 2.5$
(see Fig. S11 in the Supplementary Material of Ref.\cite{IceCube:2023ame}),  which could hint against the expected behavior from the $\pi_0$ model. 
Rather, that feature is just one of the distinctive behaviors predicted by the KRA$_\gamma$ models. However, this result is not yet statistically significant to be taken as a robust finding.}
In fact, 
the original $\pi_0$ model assumes a featureless spectrum with spectral index $- 2.7$ up to several PeVs in the whole Galaxy 
then failing to reproduce the hardening founds in the Fermi-LAT data above 10 GeV in the inner GP for the corresponding $\gamma$-ray Galactic diffuse emission ($\gamma$GDE) \cite{Fermi-LAT:2012edv,Fermi-LAT:2016zaq}. 
That is not the case for the KRA$_\gamma$ models which were originally built to reproduce that feature invoking a progressive hardening of the diffusion coefficient rigidity dependence, hence of the propagated CR spectrum, towards the Galactic Centre (GC) \cite{Gaggero:2014xla}.
As a consequence, the $\nu$GDE and $\gamma$GDE predicted by the KRA$_\gamma$ models at very high energies are significantly more peaked and harder in that region with respect to conventional models \footnote{See also Refs.~\cite{Pagliaroli_2016} and~\cite{Lipari:2018gzn} for an analytical and more phenomenological implementation of the same scenario.} including the $\pi_0$ one. This is crucial to allow these models to consistently match Fermi-LAT and the very high $\nu$GDE and $\gamma$GDE data up to the PeV from the GP including the GC Ridge \cite{Gaggero:2017jts}.

On the other hand, the KRA$_\gamma^{5(50)}$ models if normalized against the Fermi-LAT data over-predict IceCube fits by a factor 1.8(2.7), {which could be due to the oversimplified CR source spectral shape assumed for these models (due to the lack of high-energy CR data at TeV-PeV energies)}. 

A valuable independent input on the 
Very High Energy (VHE) spectral energy distribution of Galactic CRs has recently been provided by a new generation of orbital experiments as well as of Air Shower (AS) and Water Cherenkov Detector Arrays (WCDAs) which besides CR spectra and composition up to several PeVs first measured the $\gamma$GDE up to the PeV allowing an indirect probe of the CR Spectral Energy Distribution (SED) well beyond our local region. 
Among them the HAWC WCDA experiment recently measured the $\gamma$GDE between 300 GeV and 100 TeV \cite{HAWC:2023wdq} in a limited region of the GP, while Tibet AS$\gamma$ \cite{TibetASgamma:2021tpz} and LHAASO \cite{LHAASO:2023gne, LHAASO:2024lnz} probed more extended and inner regions up to the PeV. 
Although Tibet and LHAASO found apparently discrepant results -- which however may be due to the different procedures they adopt to mask resolved sources (see e.g. \cite{He:2025oys} and the Discussion below) -- both experiments agreed claiming to have found in the inner GP a $\gamma$GDE larger than expected if the Galactic CR flux would everywhere be the same as the local one. 
Since at such high energies the $\gamma$GDE is expected to be dominated by hadronic emission, those findings may have relevant implications for the interpretation of IceCube results as well.  

A significant update of the KRA$_\gamma$ models (as well as of the corresponding conventional models) was carried out in Refs.~\cite{Luque:2022buq,DelaTorreLuque:2022ats}, where the injection spectrum was re-evaluated using the most recent CR data available as well as Fermi-LAT data.
Two alternative source spectral shapes (Min and Max) were considered for the CR protons and helium nuclei in order to bracket the uncertainty in the PeV region, where the different datasets provided by KASCADE \cite{Apel:2013uni,Apel:2012tda}
differ significantly from those by
IceTop~\cite{IceCube:2019hmk} and, more recently, by LHAASO~\cite{LHAASO:2025byy}. 
The predictions of those models were then compared with Tibet AS$\gamma$ \cite{TibetASgamma:2021tpz} and preliminary LHAASO results \cite{Zhao:2021dqj} taken in the sky window $25^\circ < l < 100^\circ$ and $\vert b \vert < 5^\circ $ \cite{Luque:2022buq}. 
While finding a general consistence between IceCube results and $\gamma$-ray data, those analysis showed that the available data did not allow discriminating between Min and Max scenarios. 

Here, we take a step further: We perform a comprehensive comparison of our models to both the GeV-TeV diffuse gamma-ray data and the very high-energy data in the TeV-PeV domain. 
As in the aforementioned studies dedicated to the LHAASO and Tibet data, our aim is not to provide the best fit of the data: Instead, once the CR injection spectra are tuned to the local charged CR data, and the propagation setup is tuned on measurements of secondary-to-primary CR ratios~\cite{Tovar-Pardo:2024cbp} (both in the conventional case and in the radial-dependent case), we directly compute our prediction for the $\gamma$-ray flux with the {\tt HERMES} code \cite{Dundovic:2021ryb}. We do not further tune any additional parameter to reproduce the very-high-energy gamma-ray data.  The same procedure is followed to obtain predictions of the diffuse VHE fluxes from models of uniform propagation (conventional models, or ``Base'' models).

After presenting (Sec.~\ref{sec:model}) our improved models and 
discussing the predicted Inverse-Compton (IC) emission at energies above 100 GeV, we perform, 
in Sec.~\ref{sec:gammas}, 
a  comparison of the predictions of our $\gamma$-optimized models in different sky windows along the Galactic plane with updated Fermi-LAT observations up to several hundred GeV.
Then, we report a comparison with the recent HAWC measurements \cite{HAWC:2023wdq} between 300 GeV and 100 TeV, as well as with Tibet-AS$\gamma$ at higher energies \cite{TibetASgamma:2021tpz}. 
In this context, we extensively discuss the role of unresolved sources.
Moreover, we compare with LHAASO-KM2A~\cite{LHAASO:2023gne} data, as well as the recently released LHAASO-WCDA data~\cite{LHAASO:2024lnz} and discuss the implications of the mask used by the LHAASO collaboration to remove the contribution of sources. 
Finally, in Sec.~\ref{sec:neutrinos} the relevant implications of our findings for neutrino astronomy will be discussed including a comparison with recent ANTARES results. 
In Sec.~\ref{sec:discussion} we discuss the implications of our results and summarize our main conclusions.

\section{The $\gamma$-optimized models: an updated KRA$_\gamma$}
\label{sec:model}

In our previous works ~\cite{Luque:2022buq, DelaTorreLuque:2022ats} 
we introduced two classes of models for the $\gamma$-ray diffuse emission: the  ``Base'' and ``$\gamma$-optimized'' models, a development of the ``Conventional'' and ``KRA$_{\gamma}$'' models.

Both scenarios are based on numerical solutions of the transport equation performed with the {\tt DRAGON2} code \cite{Evoli:2016xgn,Evoli2018jcap}, a widely used public tool to compute the energy and spatial distribution of charged CR data in any point of the Galaxy\footnote{\url{https://github.com/cosmicrays/}}.
In both models, the normalization of the CR fluxes is set by local data, and the local transport properties are tuned on secondary-to-primary ratios. For more details, we refer the reader to Appendix~\ref{sec:params}.

The key difference is that, while the Base model is characterized by homogeneous diffusion along the Galactic plane (as often assumed in large-scale propagation setups), the $\gamma$-optimized model features a radial-dependent diffusion setup. 
The latter is strongly suggested by a large number of analyses of Fermi-LAT data performed over more than a decade, adopting a variety of statistical methods \cite{Gaggero:2014xla, Gaggero:2015xza, Fermi-LAT:2016zaq, Yang2016prd, Lipari:2018gzn, Pothast:2018bvh}. The remarkable phenomenological consequences of these models in the context of multi-messenger multi-TeV astronomy -- both regarding the whole Galactic plane and specific regions -- have also been pointed out in a variety of papers \cite{Gaggero:2017jts,Marinelli:2018lzs,Marinelli:2019tbu,Ventura:2021xyd}. 



The $\gamma$-optimized models provide an improved treatment upon KRA$_\gamma$ models first introduced in Ref.~\cite{Gaggero:2015xza}  
and feature updated gas maps, more accurate CR propagation setup (as detailed in \cite{Luque:2022buq}), and improved tuning on a wide range of locally measured charged CR data in the whole energy range available, from the GeV to the PeV domain.

In the following, we provide a further improvement of these models and a comprehensive comparison to all available $\gamma$-ray data from the GeV to the PeV domain \footnote{The FITS files of the updated models can be found at \url{https://github.com/tospines/Gamma-variable_High-resolution}}.

\subsection{CR nuclei source spectra: the Min and Max setups} 
\label{sec:CR_nuclei}

The considerable precision recently reached by space-based CR experiments such as PAMELA \cite{Adriani:2014pza} AMS-02 \cite{AMS:2014xys, AMS:2015tnn}, DAMPE \cite{DAMPE:2019gys,Alemanno:2021gpb}, CREAM/ISS-CREAM \cite{Ahn_2009, CREAM2017}, ATIC-2 \cite{ATIC2009}, CALET \cite{CALET:2019bmh} or NUCLEON \cite{NUCLEON2018}, 
has significantly improved  our knowledge of the composition and spectral energy distributions of primary and secondary CRs up to $\sim 100$ TeV, allowing to develop better models of particle acceleration and Galactic CR transport (see {\em e.g.} Ref.~\citep{Serpico:2018lkb}). 
However, at higher energies, these direct measurements become more challenging, and -- due to limited statistics -- only indirect measurements are available, obtained by the analysis of atmospheric showers. 
These high-energy detectors, capable to reach energies of up to $\sim 10^9$~GeV, are mostly able to measure CR protons and helium and suffer from severe uncertainties from the reconstruction of the shower at different heights. 

As a result of these complementary experimental efforts and these challenges, several features are being revealed at GeV-TeV energies showing a more complex structure with respect to a simple power-law behaviour. However, at even higher energies, the ``CR knee'' region is still subject to important uncertainties that may be hiding potential features in the spectra.

In our models, the injection of CRs is assumed to follow a broken power-law behaviour, a functional form that allows to correctly reproduce the spectral features obserbed by the aforementioned CR experiments (see for instance Sec. 3 of Ref.~\cite{Luque:2022buq}). The normalization of the spectrum is spatial dependent and follows the spatial distribution of supernova remnants as modeled in \cite{Ferriere:2001rg, Ferriere2007aa}. The slopes are instead considered uniform, since SNRs (and other classes of Galactic accelerators) are typically expected to provide a similar injection independently of their position in the Galaxy. 


In the following we do not account for CR species heavier than helium since, under reasonable conditions, their contribution to the $\gamma$-ray emission is subdominant (less than 10\%) with respect to that due to protons plus helium~\cite{Breuhaus:2022epo}  


One of the main discrepancies in the very-high-energy domain ($E \gtrsim 100$~TeV) is the significant difference between the proton spectrum measured by KASCADE \cite{Apel:2013uni} and KASCADE-Grande \cite{Apel:2012tda}, and that measured by IceTop \cite{IceCube:2019hmk} and, more recently, by LHAASO \cite{LHAASO:2025byy}
air-shower experiments on the other hand. This discrepancy  
turns into a factor of (at least) a few uncertainty in the predicted TeV-PeV diffuse $\gamma$-ray and neutrino emissions. We consider here two alternative setups, one reproducing KASCADE + KASCADE-Grande, as well as CALET \cite{CALET:2019bmh,Akaike:2024dvl} and DAMPE \cite{DAMPE:2019gys} at lower energies (what we call ``Min'' setup) and another one following the measurements by IceTop (the ``Max'' setup).
The local (propagated) spectra (fitted to data) of CR proton and helium for the Min and Max setups are shown in Fig.~\ref{fig:CRnuclei} (Appendix~\ref{appendix:CR}), compared to existent CR data sets, covering around nine orders of magnitude in energy. Complete details are provided in \cite{Luque:2022buq} (see also Ref.~\cite{DelaTorreLuque:2022ats}).
We note that having more and more precise measurements of the $\gamma$-ray diffuse emission at different regions of the Galaxy would allow us to infer the distribution of CRs, or, at least, to have valuable complementary information to local CR data.

\subsection{CR electron and Inverse-Compton emission}
\label{sec:CRe}

\begin{figure}[!t]
\centering
\includegraphics[width=0.495\textwidth]{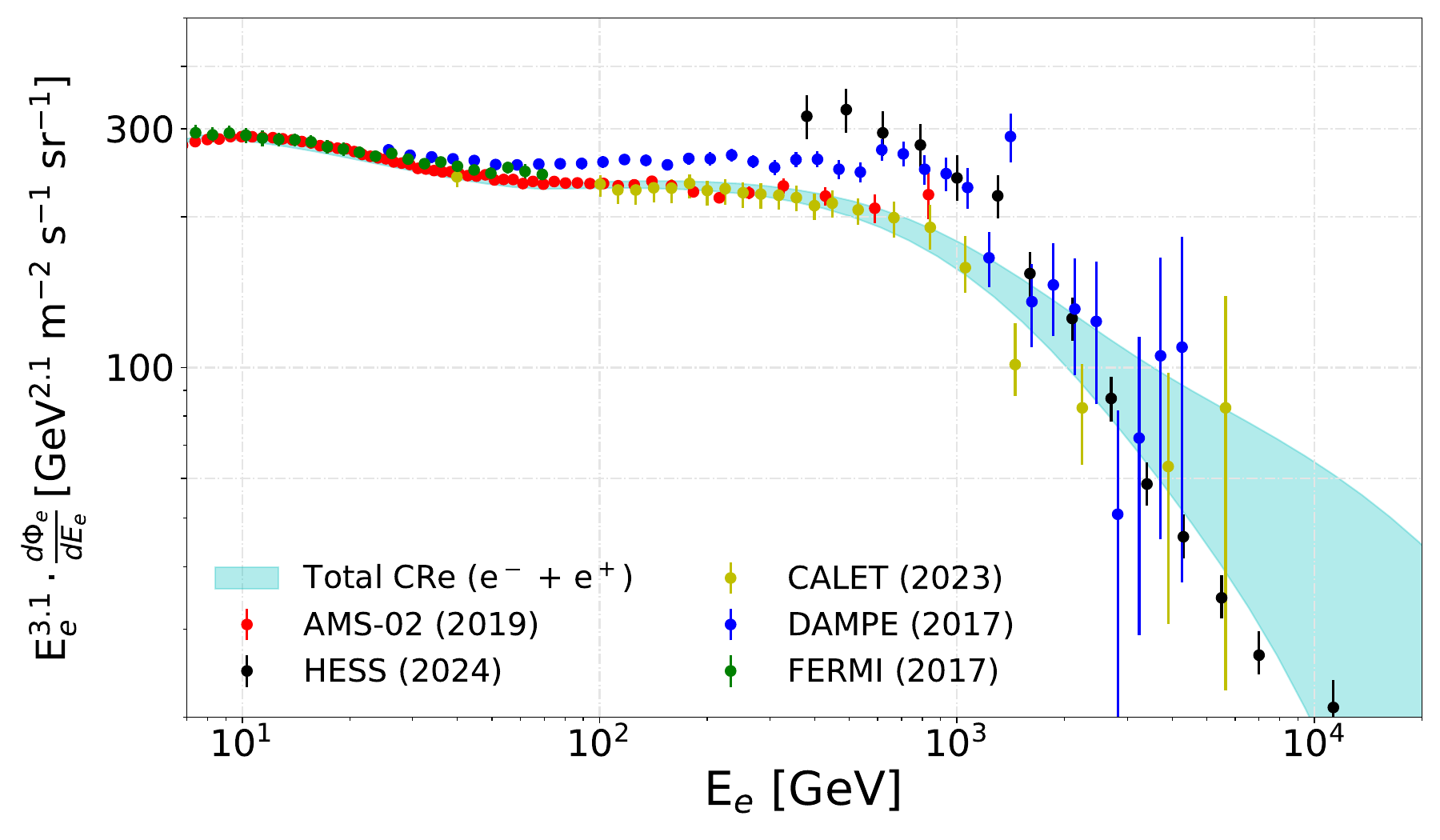}
\includegraphics[width=0.495\textwidth]{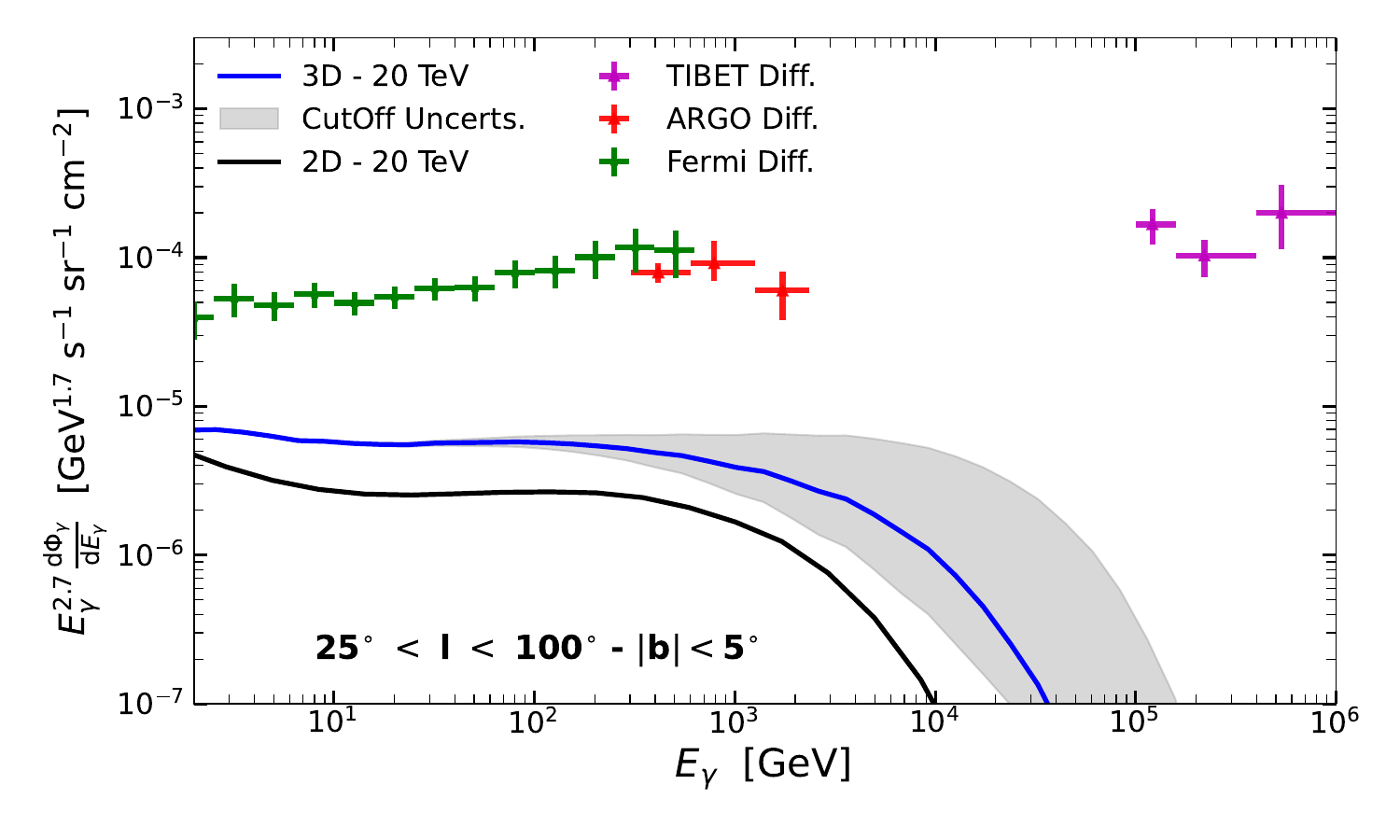} 
\caption{\textbf{Left panel:} Local CRe spectrum estimated in our model, compared to available data. The uncertainty band represent the uncertainty in the energy cut-off of the injection of these particles. \textbf{Right panel:} Predicted IC emission up to PeV energies (with energy cutoff at $20$~TeV), compared to available measurements in a region covering the Galactic plane. The blue line indicates the estimated emission when adopting the spiral arms in the CRe source density distribution, while the black line indicates the estimated emission when adopting a cylindrically symmetric 2D source distribution. }
\label{fig:EleUncerts}
\end{figure}

The leptonic component of the CR sea produces $\gamma$-rays via bremsstrahlung and IC scattering.
In order to estimate these contributions, we need to model the spatial distribution and spectrum of CR electrons (and positrons) in the Galaxy. 
We use {\tt DRAGON2} to compute the steady state solution of the diffuse electron/positron distribution, based on the local electron and positron spectra measured from various detectors~\cite{HESS:2018pbp, Akaike:2024dvl, Abdollahi_2017, ATIC2009, AMS:2014xys, DAMPE_e}. 

A key aspect in this context is the fact that the high-energy positron spectrum (above a few tens of GeV) may be dominated by the emission from pulsars, which inject equal amounts electrons and positrons. We describe this component as a broken power-law with a cut-off, tuned to the AMS-02 local positron data. 
Therefore, as far as the total lepton spectrum is concerned, we perform our modeling under the assumption that two classes of sources are at work: SNRs (that accelerate electrons), and pulsars (that accelerate electrons and positrons in equal amounts). The injection spectrum of SNRs in tuned on the electron spectrum, while the pulsar contribution is tuned on the positron component. 

We show our results concerning the leptonic component of the CR sea in Fig. \ref{fig:EleUncerts}. In order to bracket the large uncertainties associated to the different normalization of different datasets (possibly due, among other effects, to energy calibration), we consider a range of rigidities for the high-energy cutoff of the injection spectrum. The range [$100$ GeV - $100$ TeV] is adequate to capture the large scatter that characterize the data reported by the most important experimental collaborations in this channel.

Another important ingredient in this computation is the distribution of sources in the Galaxy, due to the fact that electrons and positrons lose energy very quickly in the high-energy domain as a consequence of to Inverse Compton and synchrotron emission. 
Here we provide a more refined model for this astrophysical ingredient with respect to the previously published models: We adopt a spatial distribution of sources that follows the spiral arm distribution of our Galaxy, based on the 4-arm model derived in Ref.~\cite{Steiman-Cameron_2010}. 
This setup is relevant only as far as the leptonic component is concerned (protons do not suffer from the aforementioned energy losses in the high-energy domain). We will refer to this scenario as the {\it 3D model}.

While both the previously considered cylindrical model and the 3D model lead to equally good fit of local CRe data, the injection spectra is harder for the case of a source distribution following the spiral arms. 
This is needed to compensate the fact that the average distance of a source with respect to the Earth is larger in the 3D model, because the Earth is located at the edge of the Sagittarius arm, resulting in severe energy losses for CRe. 
As a consequence, the average IC emission in the plane of the Galaxy increases by a factor of a few in the 3D model we consider here. 
This is shown in the right panel of Figure~\ref{fig:EleUncerts}, where we compare a variety of measurements in the Galactic region at $|b| < 5^{\circ}$ and $25^{\circ} < l < 100^{\circ}$ with the expected IC emission predicted from with the 3D source distribution (blue line) and from a cylindrical distribution of sources (the Ferriere distribution, also used for CR nuclei, as discussed above). We display here, around the prediction from the 3D model, the uncertainty band due to the variations in the cutoff from $1$ to $100$~TeV. The lines describe the predicted IC emission assuming the cutoff at $20$~TeV. As seen, the difference between both estimations is above a factor of $2$ at $10$~TeV, but the total emission is always lower than a $10\%$ of the measured $\gamma$-ray emission at that region of the Galaxy and totally negligible above a few tens of TeV.

We note that a potential limitation of our modeling is that we are assuming a smooth distribution of electron sources, while a Monte Carlo approach that accounts for a stochastic distribution of sources could be more accurate at these energies (see for instance \cite{Evoli:2021ugn}). 


Finally, we note that we disagree significantly with Ref.~\cite{Marinos:2024rcg}, where the authors obtain a dominant IC emission at TeV energies. This can be due to the fact that their median distribution of local electrons appears to be significantly higher than the one that we fit to local data. In fact, the new H.E.S.S. data~\cite{Aharonian_2024} seems to be in significant tension with their estimated $68\%$ confidence interval. We also show that our estimations are compatible with Fermi data above and below the Galactic plane, where the IC emission becomes more significant, in the right panels of Figs.~\ref{fig:AbovePlane} and~\ref{fig:AbovePlane_App}.
The study of synchrotron emission in these regions of the Galaxy will be fundamental to constrain better the CRe spectrum. Unfortunately, to constrain the cutoff energy, we  need measurements at frequencies above, at least $10^4$~GHz, where data is scarce and there is also an important free-free emission. In addition, uncertainties in the magnetic field intensity prevent us from obtaining robust conclusions. Future CTA \cite{CTAConsortium:2023tdz} data on the CRe local flux and anisotropies will certainly help us understand the role of nearby sources and the  IC emission in the Galaxy.

\section{Predicted $\gamma$-ray Galactic emission from 10 GeV to 1 PeV}
\label{sec:gammas}

In this section, we compare the prediction of our $\gamma$-optimized models with the measurements taken by several $\gamma$-ray experiments over a wide range of energies. %
The spectrum of the $\gamma$GDE measured by this experiment provides an indispensable lever arm to model the neutrino and $\gamma$-ray diffuse emissions at larger energies.



To determine the spectrum of $\gamma$-rays produced by the scattering of CR protons and helium onto the interstellar hydrogen and helium gas we use the {\tt AAFRAG} packgage \cite{Kachelriess:2019ifk,Kachelriess:2022khq} which, respect to other parameterizations, minimizes the computed emission and is the most updated. The uncertainty associated to that choice has been discussed in Ref.~\cite{DelaTorreLuque:2023usg} (see also Refs.~\cite{Vecchiotti:2024kkz, Schwefer_2023}).

All the spectra and angular distributions of the $\gamma$, as well as of the $\nu$, interstellar emission shown below are obtained with the {\tt HERMES} code \cite{Dundovic:2021ryb} which also accounts for opacity due $\gamma - \gamma$ scattering onto the CMB as shown in Ref.~\cite{Luque:2022buq}. Details about the interstellar gas distribution used in this work can be found in that paper. We remark that the predictions of the hadronic diffuse emission that we show in the following sections were already available in Refs.~\cite{Luque:2022buq, DelaTorreLuque:2022ats}. Here, we perform a detailed and comprehensive comparison with all the recently released high-energy data sets without any further tuning on the aforementioned prediction (as far as the hadronic component is concerned). 

\subsection{Comparison with Fermi-LAT data}
\label{sec:Fermi}

The Fermi-LAT collaboration has provided the most detailed and accurate information on the $\gamma$-ray sky over a wide energy range from from $\sim0.1$~GeV to from $\sim1$~TeV, with angular resolution reaching $0.2^{\circ}$ in the range from $3$ to $300$~GeV.  These data are currently the most valuable tool to build templates aimed at studying the very-high-energy emission in multi-messenger context.
This $\gamma$-ray map allowed to infer the gradual change in the slope of the hadronic component towards the center of the Galaxy that we discussed in the previous Sections~\cite{Gaggero:2014xla,Fermi-LAT:2016zaq}. 
Such trend has been confirmed by a variety of studies, and was shown to persist above few tens of GeV \cite{Pothast:2018bvh}.


\begin{figure}[t]
\centering
\includegraphics[width=0.495\textwidth]{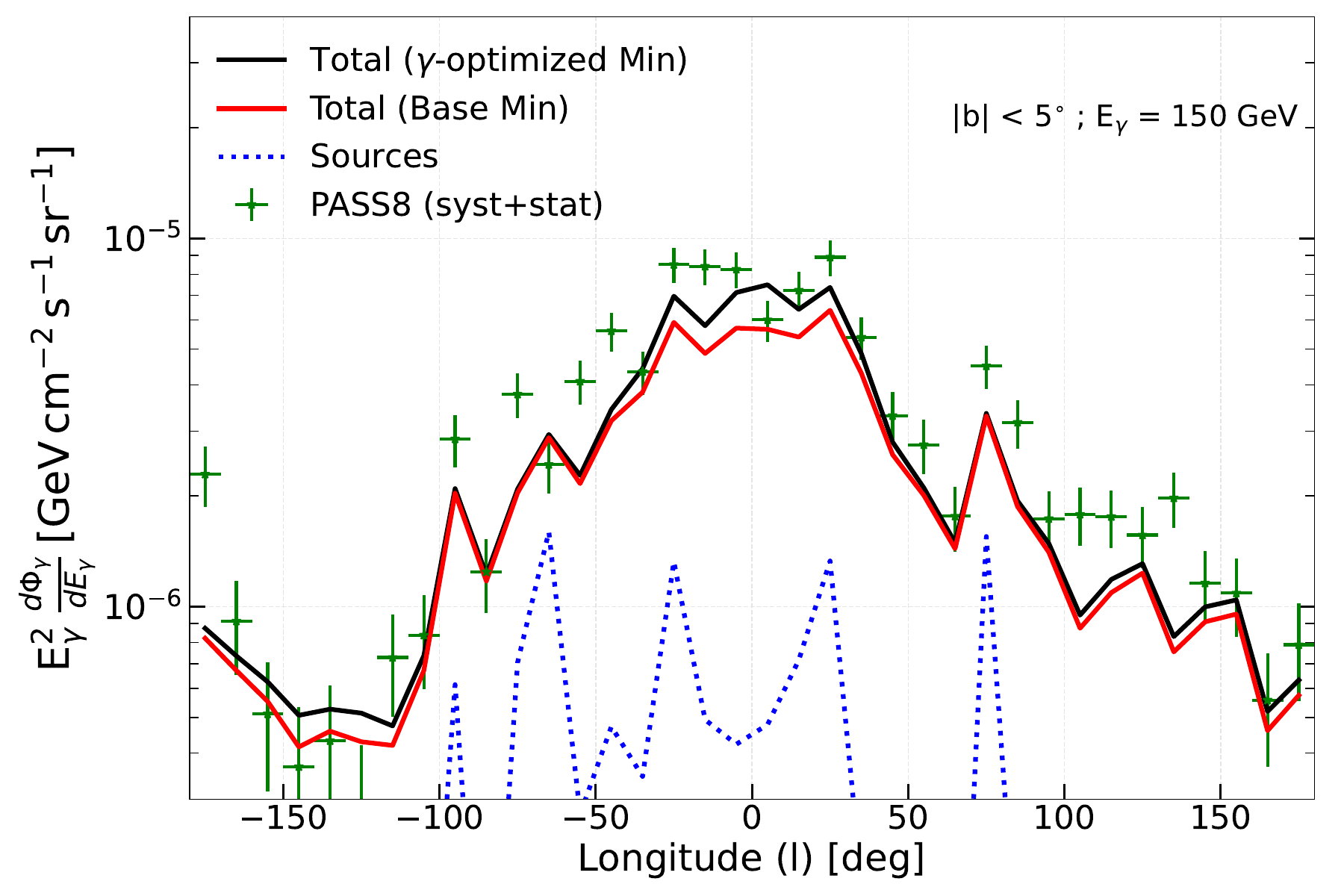}
\includegraphics[width=0.495\textwidth]{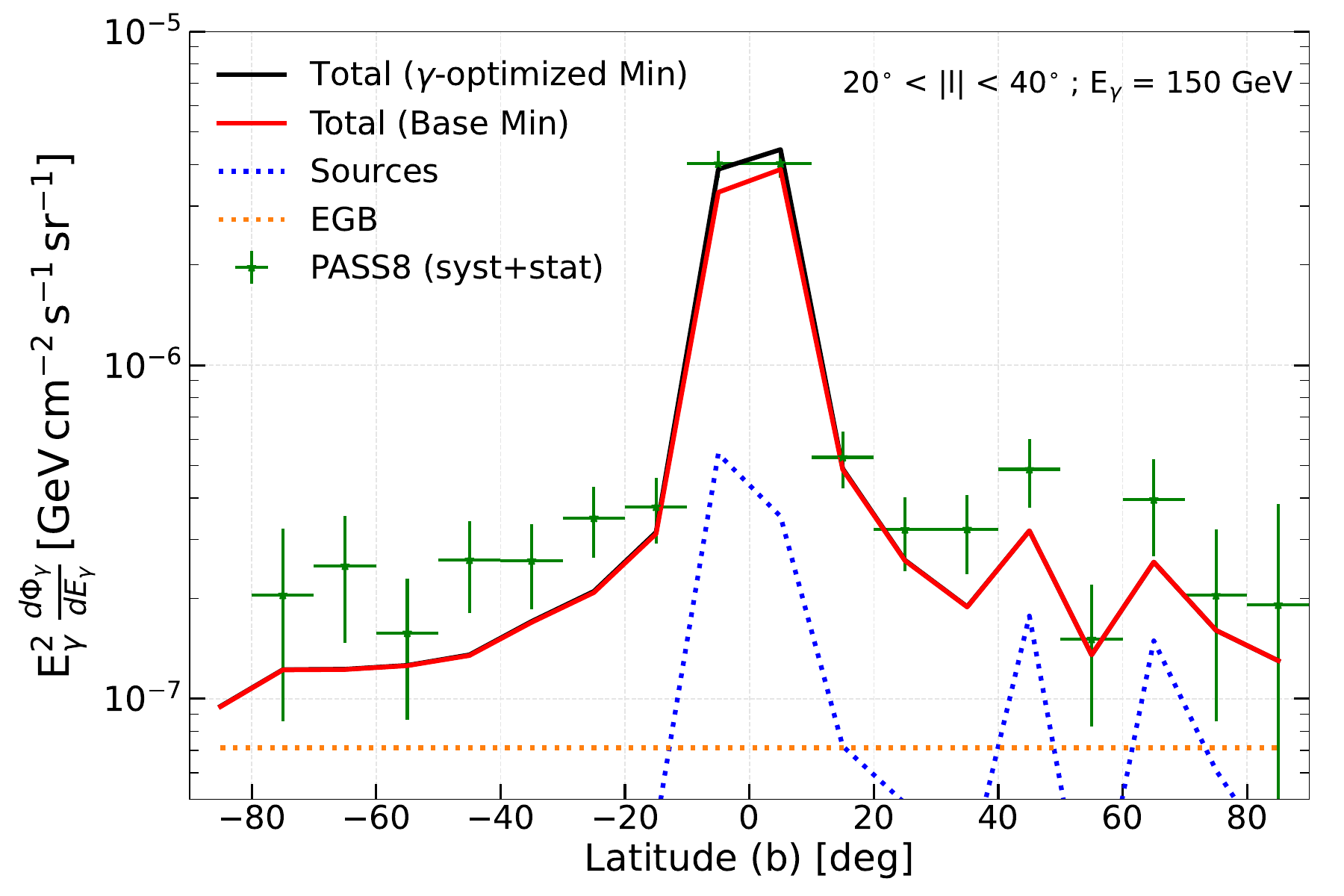}
\caption{Longitude (left) and latitude (right) profiles of the total $\gamma$-ray emission as measured by Fermi-LAT (as green data-points), in comparison to the predicted total emission from the $\gamma$-optimized (Black line) and Base models (Red line). 
The predicted total emission is the sum of diffuse emission from CRs (our models), source emission (from the 4FGL-DR2 catalog), extragalactic (isotropic) emission and the estimated flux from unresolved point-like sources. The contribution from point-like sources (Sources - blue dotted line) and the isotropic, extra-galactic emission (EGB, yellow dotted line) are also shown. A similar figure can be found in Appendix~\ref{appendix:Fermi_Contrib}, Fig.~\ref{fig:Fermi_profs2} for different sky regions. }
\label{fig:Fermi_profs}
\end{figure}

In this Section we present a comparison between our models and Fermi-LAT data. This data is extracted as detailed in Ref.~\cite{Luque:2022buq}, where we refer the reader for detailed information on the data reduction.
In more detail, we test our prediction taking into account:

\begin{itemize}

    \item The sum of the {\it truly diffuse} gamma-ray emission ($\pi^0$ emission, Inverse Compton and bremsshtrahung) obtained with the {\tt HERMES} code as detailed in the previous Section.

    \item The emission from the point sources reported and modeled in the 4FGL-DR2~\cite{Fermi4FGL} catalog

    \item The isotropic background as provided by the Fermi-LAT collaboration\footnote{We adopt here the template that refers to CLEAN data: (iso\_P8R3\_CLEAN\_V3\_v1) from \url{https://fermi.gsfc.nasa.gov/ssc/data/access/lat/BackgroundModels.html}}. This component is actually a sum of the ``true'' extragalactic diffuse flux and the residual mis-classified CRs. 

    \item A model for the unresolved source component. This contribution is modeled by population synthesis, following a four-arm spiral distribution described in \cite{SteppaEgberts2020}. More details on this can be found in our previous paper \cite{Luque:2022buq}.

\end{itemize}

The dataset consists in more than 12 years of Fermi-LAT data, analyzed as explained in Ref.~\cite{Luque:2022buq}, to which we refer the reader for full details in the data extraction.

The first comparison we present concerns the Galactic longitude and latitude profiles, displayed in Fig.~\ref{fig:Fermi_profs}. The ``Total'' lines represented in this panel correspond to the sum of the four components listed above, for both our updated ``Base'' and ``$\gamma$-optimized'' (i.e. featuring inhomogeneous diffusion) model. {In these analysis we avoid comparing the emission at high latitudes in regions where the Fermi Bubbles can play an important role.
Note that we are using the Min setup as a reference in these figures because at GeV energies there is no difference between the Min and Max setups.}

%
 Overall, we point out a good agreement between the relevant features exhibited by the data and our models, especially as far as the longitude profile is concerned. This is a consequence of the detailed gas model implemented in {\tt HERMES}, that allows to trace the morphology of dominant emission from $\pi^0$ decay with remarkable accuracy. We notice that the difference between the $\gamma$-optimized and Base  model becomes relevant in the inner $|l|\lesssim 70^{\circ}$ and $|b|\lesssim10^{\circ}$ , as expected given the progressive hardening towards the center that characterized the former class of models.

Moreover, we see that emission from sources can be very important in certain regions of the Galaxy. In turn, we observe that emission at high latitudes is dominated by the isotropic (Extra-Galactic) background, with a non-negligible  subdominant contribution from the diffuse emission. 
We show a similar figure for other regions of the sky (showing a similar agreement to data) in Fig.\ref{fig:Fermi_profs2} of App.~\ref{appendix:Fermi_Contrib}, and other similar comparisons can be found in Ref.~\cite{DelaTorreLuque:2022ats}. 

\begin{figure}[!t]
\centering
\includegraphics[width=0.48\textwidth]{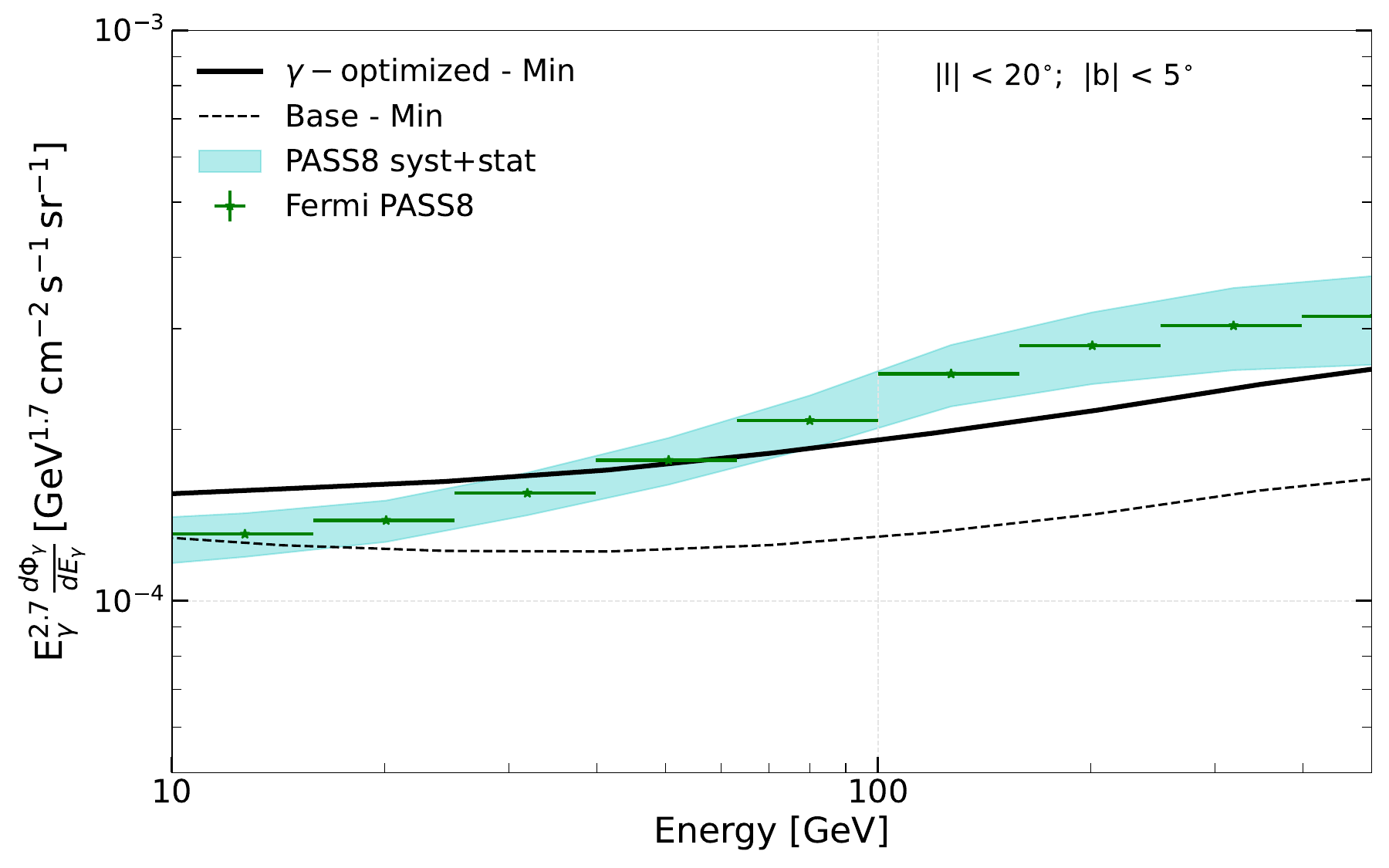} 
\hspace{0.1cm}
\includegraphics[width=0.48\textwidth]{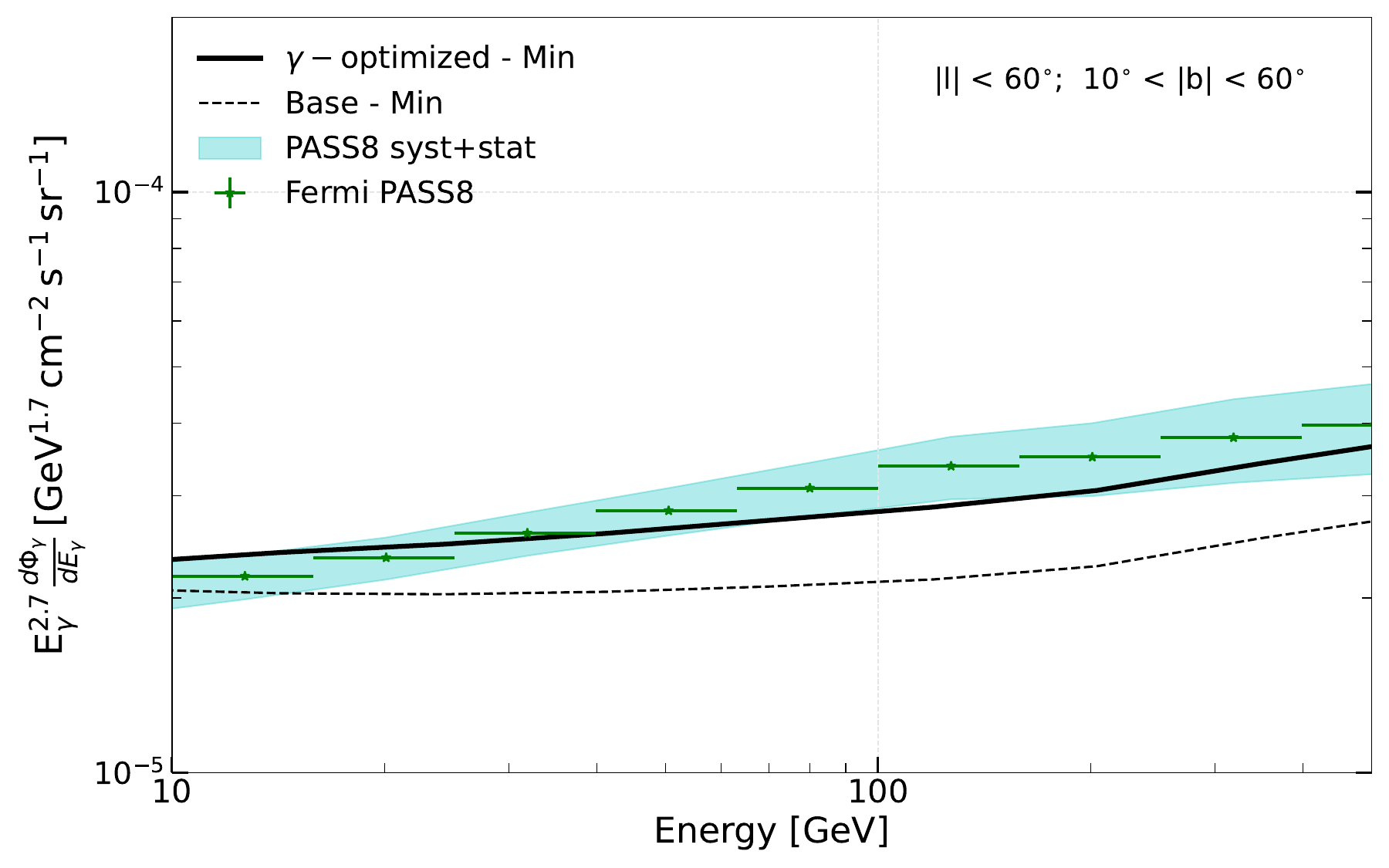}

\includegraphics[width=0.48\textwidth]{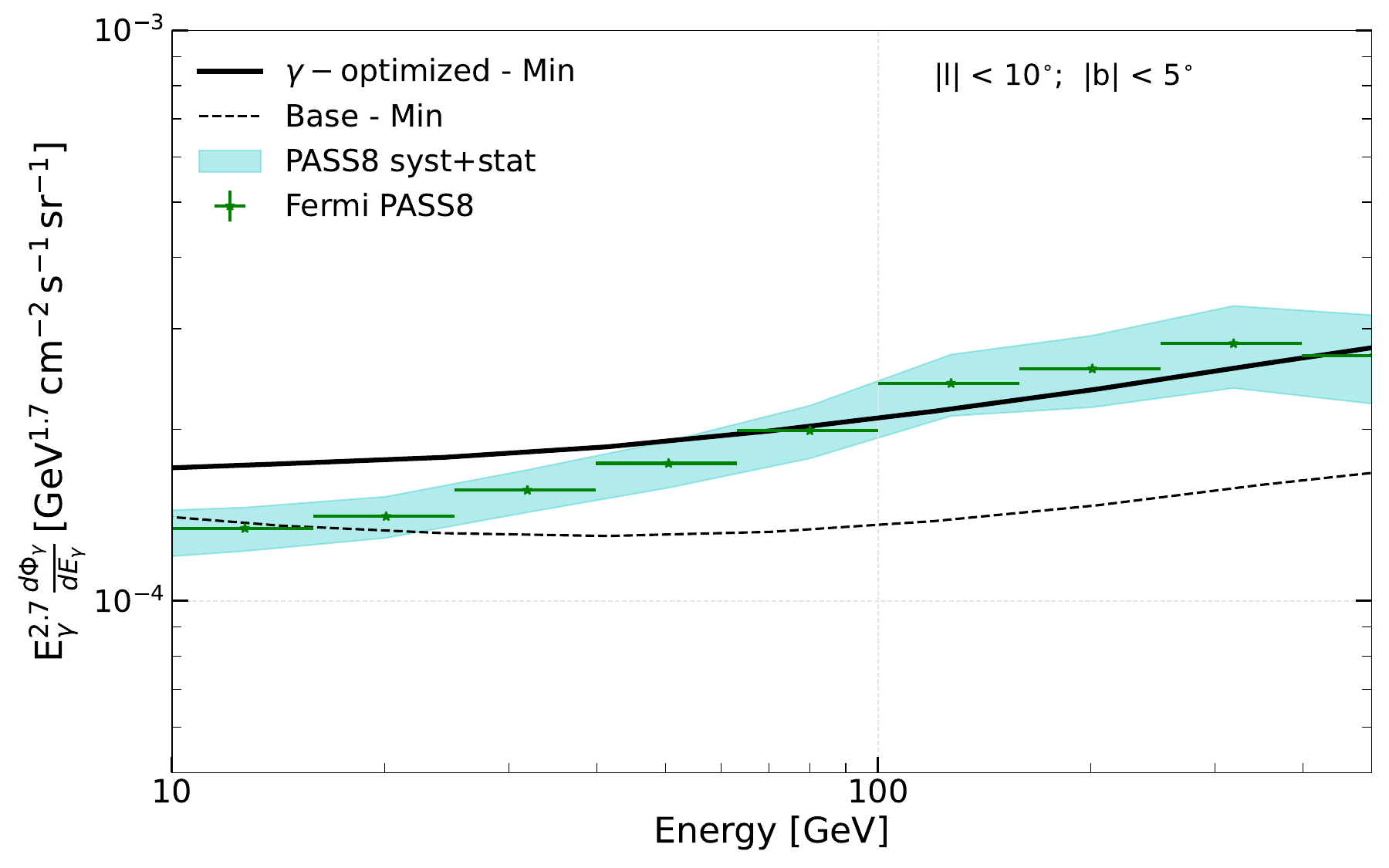}
\hspace{0.1cm}
\includegraphics[width=0.48\textwidth]{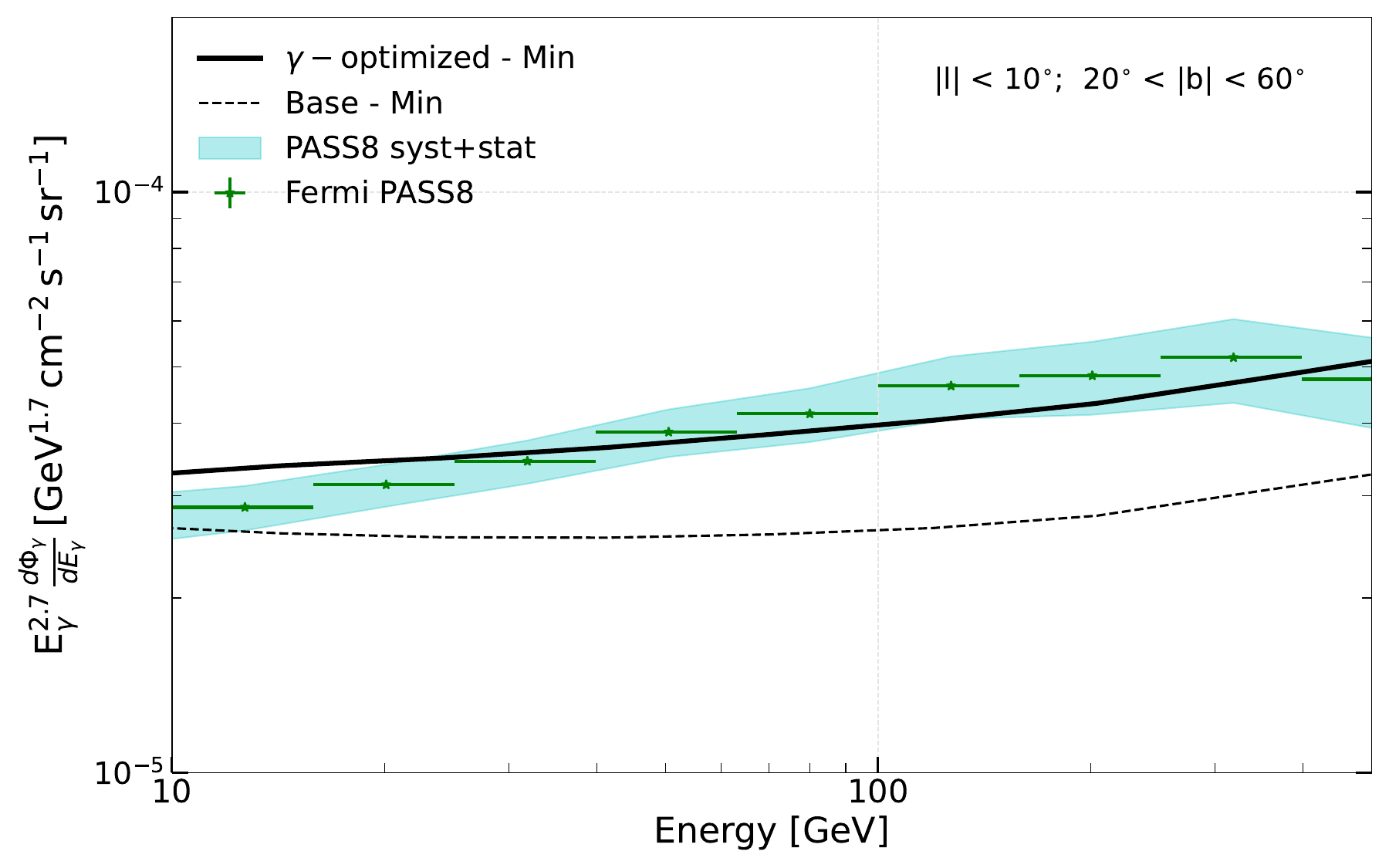} 
\caption{Comparison of Fermi-LAT measurements in different regions of the Galaxy with the total $\gamma$-ray emission considering the $\gamma$-optimized (solid lines) and Base (dashed lines). For both models, the Min setup is used, summed to the source emission (from the 4FGL-DR2 catalog), isotropic background and point-like unresolved source estimated emission for Fermi-LAT. The blue band on top of Fermi-LAT data represents the statistical + systematic uncertainty of these measurements.}
\label{fig:AbovePlane}
\end{figure}

As we see from these comparisons, the inhomogeneous diffusion setup with a hardening towards the GC improves the agreement of our predictions with Fermi-LAT observations, as highlighted in the aforementioned statistical analyses performed over a decade.

More evidence favoring this scenario is shown in Fig.~\ref{fig:AbovePlane}, where we report a {\it spectral} comparison of the predictions for the $\gamma$-optimized and Base models -- both in their Min setup -- with the Fermi-LAT data in four different sky windows. 
The cyan bands around the data represent the systematic uncertainties from the effective area (plus statistical uncertainty, that is usually negligible in comparison). 
The two left panels focus on a region that contain the GC, where hadronic emission is significantly dominant, while the right panels focus on regions above the GP, where the IC contribution becomes more relevant. As we see here, the difference between both models becomes very important in comparison with the Fermi-LAT uncertainties in the measurement.
%
In Appendix~\ref{appendix:Fermi_Contrib} Fig.~\ref{fig:AbovePlane_App}, we show the same figure as Fig.~\ref{fig:AbovePlane} but detailing the different components of the $\gamma$-ray emission
As it can be seen from Fig.~\ref{fig:AbovePlane_App} (App.~\ref{appendix:Fermi_Contrib}), while bremsstrahlung contribution is always very subdominant (contributing, at most, at the percent level), the IC contribution constitutes around a $10\%$ of the total emission in the Galactic plane and larger than a $\sim25\%$ out of the plane (especially, this can be seen in the bottom-right panel, where the IC contribution reaches around a $35\%$ of the total flux). 

We also note that the contribution from unresolved sources, although increasing its importance with energy, remains very low for Fermi-LAT even at high latitudes. However, our estimation of the contribution of unresolved sources to Fermi-LAT observations is only accurate below $\sim100$~GeV. This is due to the lower effective area above these energies and the fact that above $\sim300$~GeV the spatial resolution of the detector is significantly poorer.

We note a remarkable agreement of our model with Fermi-LAT observations above few tens of GeV in different zones of the Galaxy (more details can be found in Refs.~\cite{Luque:2022buq, DelaTorreLuque:2022ats}). 
We also note that our model is based on a different, more physical, approach than the Fermi-LAT diffuse template: In fact, the latter requires the independent tuning of several spatial templates in each energy bin and the addition of dedicated patches. 
However, the accuracy of our modeling of Fermi-LAT data is enough for the purpose of the present paper, given the large uncertainties in the TeV-PeV data. 

\subsection{Comparison with HAWC results}
\label{sec:HAWC_comp}

\begin{figure}[t]
\centering
\includegraphics[width=0.6\textwidth]
{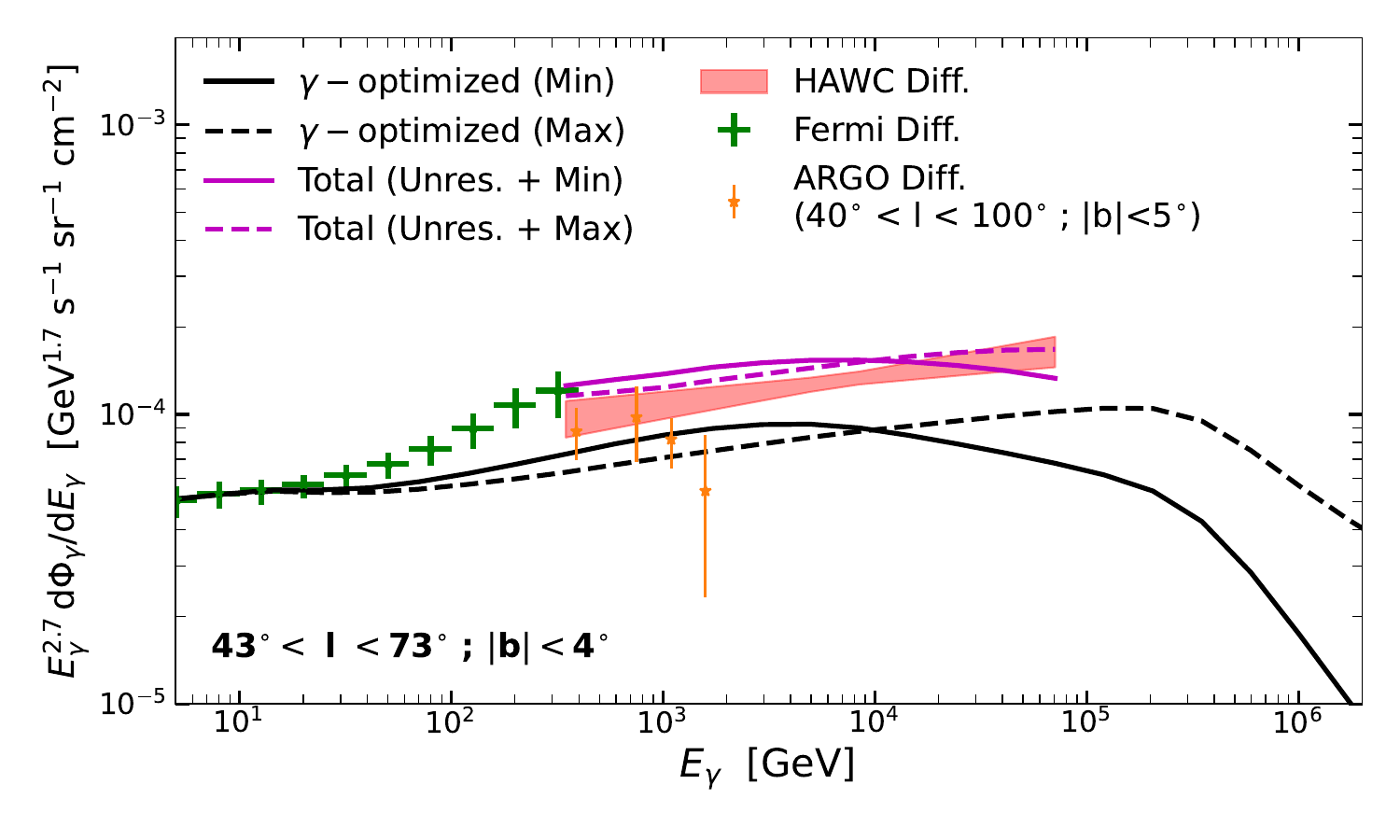}
\caption{Comparison of the HAWC $\gamma$DGE observations (as a red band, that includes statistical errors) at $43^{\circ} < l < 73^{\circ}$ and $|b|<4^{\circ}$ with the $\gamma$-optimized Min (solid lines) and Max (dashed lines) models. Fermi-LAT (in green; with errorbars representing the cuadratic sum of statistical and systematic uncertainty on the effective area) and ARGO-YBJ~\cite{ARGO-YBJ:2015cpa} data (in orange) are also shown. For completeness, we add as magenta lines the sum the expected contribution from unresolved sources (estimated in Ref.~\cite{Eckner_Limits}) to the truly diffuse emission from the $\gamma$-optimized models.}
\label{fig:HAWC_Comp}
\end{figure}

The High-Altitude Water Cherenkov Observatory (HAWC) has recently provided measurements of the $\gamma$DGE~\cite{HAWC:2023wdq} in different regions of the the Galactic plane (for $l \in [43^{\circ}, 73^{\circ}]$) at $|b|<2^{\circ}$ and $|b|<4^{\circ}$ (shown in Fig.\ref{fig:HAWC_Comp}).
In addition to these averaged fluxes, they provided the longitudinal (from $|b|<2^{\circ}$ to $|b|<4^{\circ}$, with $3^{\circ}$ wide longitude bins) and latitudinal emission profiles (from $b=-4.5^{\circ}$ to $b=4.5^{\circ}$ in 9 latitude bins $1^{\circ}$ wide, at $43^{\circ} < l < 73^{\circ}$) (shown in the Fig.~\ref{fig:Profs_HAWC}) in Appendix~\ref{appendix:HE_comparisons}. 
The HAWC diffuse analysis \cite{HAWC:2023wdq} is based on a template-fitting technique, and adopts an older version of the Base model (Base$^5$)~\cite{Gaggero:2015xza} as a spatial template for the $\gamma$GDE. The emission of point-like and extended sources from H.E.S.S. \cite{HESS:2018pbp} and third HAWC \cite{HAWC:2020hrt} catalogs was subtracted.  
The $\gamma$GDE spectrum in the region $43^{\circ} < l < 73^{\circ}$, 
 $|b|< 4^{\circ}$ was fitted with a power-law with index $-2.60 \pm 0.03$, finding a  normalization a factor two higher than the Base$^5$ model~\cite{Gaggero:2015xza} prediction. The systematic uncertainties of the measurement are not included in the analysis.

 In Fig.~\ref{fig:HAWC_Comp}, we compare the spectrum measured by HAWC (as a salmon red band) at $43^{\circ} < l < 73^{\circ}$ and $|b|<4^{\circ}$ to our $\gamma$-optimized Min and Max models (black lines). In this figure, we also include Fermi-LAT diffuse data (i.e. the emission after subtracting point-like sources and the EGB) and ARGO data \cite{ARGO-YBJ:2015cpa} (available in a slightly different region, at $40^{\circ} < l < 100^{\circ}$ and $|b|<5^{\circ}$). We did not apply any mask here, which is expected to only affect our result with a slight reduction of the predicted diffuse flux, given the very small portion of the Galactic plane that is masked in the HAWC analysis. We notice $\lesssim 2$ factor difference between the HAWC measurement to the predicted emission from the $\gamma$-optimized model (and similarly for the most inner region at $|b|<2^{\circ}$ in Appendix~\ref{appendix:HE_comparisons}). This is expected because the contribution from unresolved sources must be significant for the HAWC detector in this region (see, e.g.~\cite{Lipari:2024pzo, Vecchiotti_2022, Eckner_Limits}).  
To show this, we also include an estimation of the total $\gamma$GDE emission (magenta lines) in this figure, which is the result of summing the truly diffuse emission from CR interactions (from the Min and Max $\gamma$-optimized models) plus the contribution from unresolved sources that was estimated in Ref.~\cite{Eckner_Limits}, where the authors showed a remarkable agreement between previous HAWC measurements and the total $\gamma$GDE. As we see, summing both contributions, which were derived in a totally independent way, lead to a reasonable match of HAWC observation (which includes only statistical errors).  
Unfortunately, given the relevance of unresolved sources here and the fact that in this region of the sky the difference between our Base and $\gamma$-optimized models is only of tens of percent (much lower than the uncertainties in the measurement), current HAWC measurements cannot provide evidence for uniform or inhomogeneous CR propagation. However, these observations clearly show that the contribution from the truly diffuse $\gamma$-ray emission (i.e. the one only coming from CR interactions) must be very relevant, and probably dominant at TeV energies. Moreover, no clear preference for the Min or Max setups can be obtained at those relatively low energies.
We note that the Base model is not shown in this figure because it differs by only $\sim10\%$ from the $\gamma$-optimized model.

In the Appendix \ref{appendix:HE_comparisons} (Fig.~\ref{fig:Profs_HAWC}) we also compare our models with the $\gamma$GDE longitude (right panel) and latitude (left panel) profiles measured by HAWC (integrated between $0.3$ and $100$ TeV). We note here that while the regions at high latitude seem to be reasonably well described by the truly diffuse emission, the regions closer to the Galactic plane (where we expect more sources~\cite{Lipari:2024pzo, SteppaEgberts2020, Vecchiotti_2022, Kaci:2024lwx}) require a significant, although still subdominant, contribution from unresolved sources.

\subsection{Comparison with Tibet-AS$\gamma$ results}
\label{sec:Tibet_comp}

\begin{figure}[t]
\centering
\includegraphics[width=0.495\textwidth]{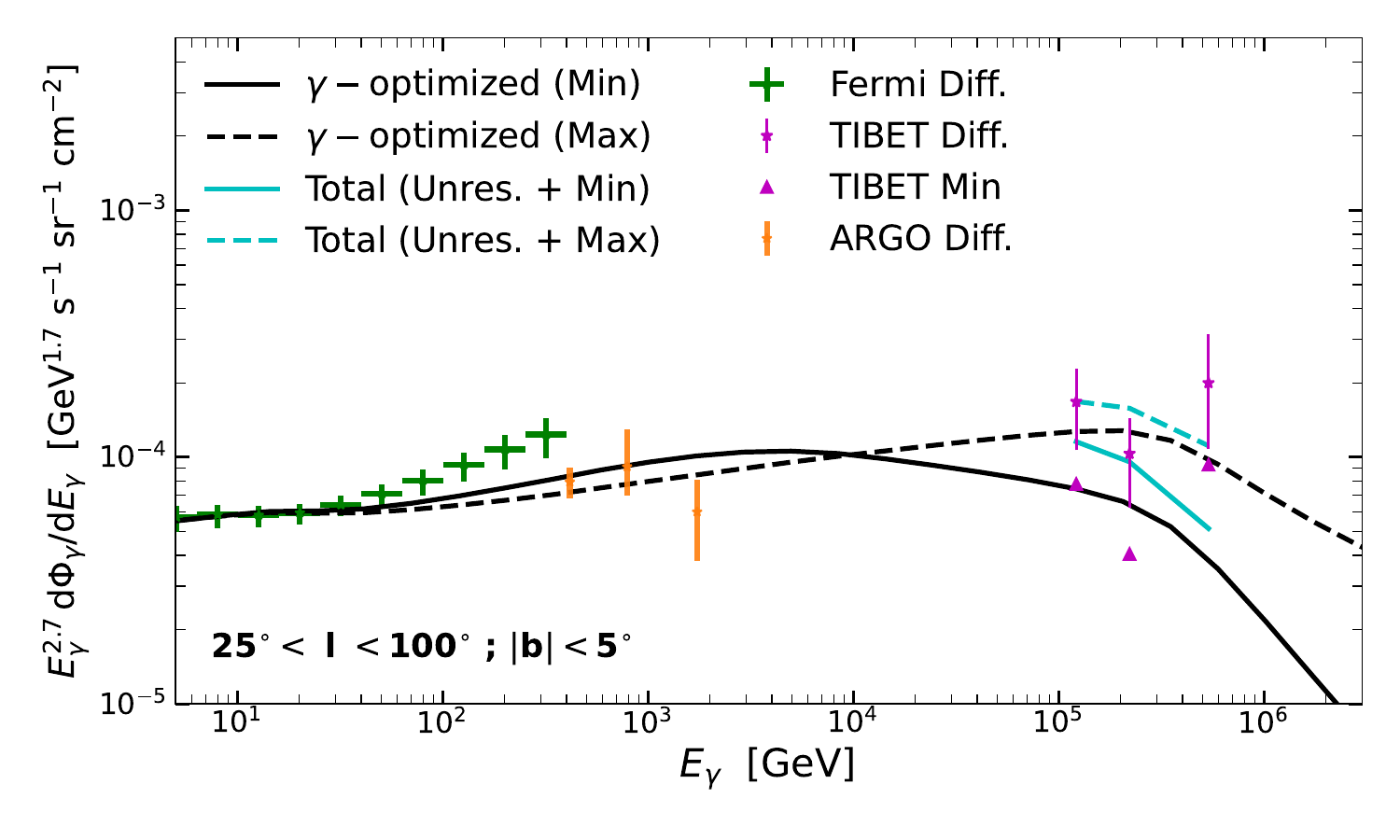}
\includegraphics[width=0.495\textwidth]{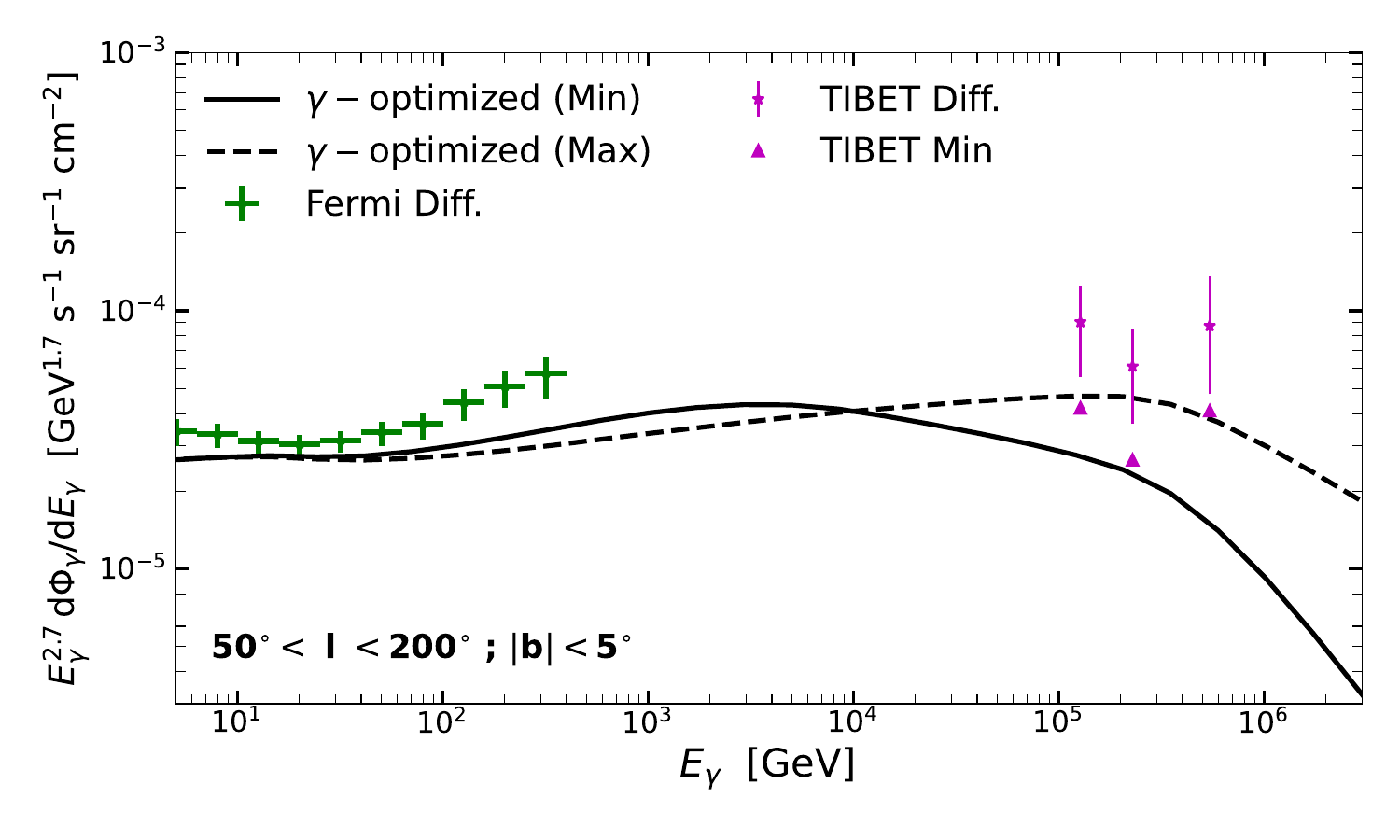}
\caption{$\gamma$GDE in the TIBET Region A (left panel) and Region B (right panel), compared to the predicted emission from the $\gamma$-optimized models Min (solid lines) and Max (dashed lines). For the left panel (TIBET Region A), we also show the total diffuse emission (as unresolved source emission, from Ref.~\cite{Eckner_Limits}, plus the truly diffuse emission from the $\gamma$-optimized models), as cyan lines. Fermi-LAT errorbars include statistical plus systematic errors (from the effective area) while TIBET measurements only include the $1\sigma$ statistical uncertainty. The magenta arrows indicate the $1\sigma$ lower flux subtracting the contribution from resolved extended sources, as studied in Ref.~\cite{Kato:2024ybi}.}
\label{fig:TIBET_Comp}
\end{figure}

The first detection of the $\gamma$GDE above $100$~GeV was achieved by the Tibet-AS$\gamma$ collaboration~\cite{TibetASgamma:2021tpz}. 
To extract the $\gamma$GDE, Tibet adopts a mask to avoid emission from the point-like sources reported in the TeVCat removing $0.5^{\circ}$ around their nominal positions. Again here, the mask only subtracts a very small portion of the GP, therefore resulting in a measurement that is expected to represent the real $\gamma$GDE in the $100 ~{\rm TeV} < E < 1 ~{\rm PeV}$ range. The Tibet collaboration provided their observation in two regions: one at $25^{\circ} < l < 100^{\circ}$, $|b| < 5^{\circ}$ (region A) and another at  $50^{\circ} < l < 200^{\circ}$, $|b| < 5^{\circ}$ (region B) -- which are shown in Fig.~\ref{fig:TIBET_Comp}. 
Remarkably, in Ref.~\cite{Kato:2024ybi}, the authors showed that none of the $23$~Tibet diffuse $\gamma$-ray events detected above $398$~TeV come from the LHAASO catalog sources detected above $100$~TeV, supporting the diffusive nature of these events. However, the fact that the emission around certain sources (especially, the Cygnus cocoon~\cite{LHAASO_Collaboration_2024}) can be more extended than what they assumed, have triggered discussions about the possible overestimation of the $\gamma$GDE measured by Tibet~\cite{Yan:2023hpt, 2021ApJ...914L...7L}. In this respect, we also remark that the highest-energy Tibet data point is estimated from only $10$~events, with a few of them coming from a region close to the Cygnus cocoon.
Interestingly, a few studies~\cite{Yan:2023hpt, 2021ApJ...914L...7L, 2023ApJ...957L...6F} have pointed to this as the origin of the discrepancy between the $\gamma$-GDE measured by LHAASO and Tibet. 
Here, we find that Tibet $\gamma$-GDE is between a few tens of percent and a bit more than a factor of $2$ larger than the flux measured by LHAASO in both, inner and outer regions, after rescaling their measurements to account for the effect applying the LHAASO mask and the slightly different regions where the experiments provided their observations (see more details in the next section).

In Fig.~\ref{fig:TIBET_Comp} we compare our $\gamma$-optimized Min and Max models with Tibet-AS$\gamma$ results. 
The $\gamma$GDE spectral observations from Fermi-LAT and ARGO-YBJ \cite{ARGO-YBJ:2015cpa} in the same region are also reported (ARGO data is only available for Region A).

Let us clarify some aspects of this comparison:
\begin{itemize}
\item Given the large energy considered here, the absorption due to scattering onto CMB and other radiation fields becomes significant and is properly taken into account in the {\tt HERMES} modeling, as described in Ref.\cite{Dundovic:2021ryb,Luque:2022buq}. 
\item We do not apply any mask to our models, since we expect that the Tibet mask would only affect our prediction at percent level. 
\item As far as unresolved sources are concerned Refs.~\cite{Lipari:2024pzo, Eckner_Limits, Kaci:2024kog} 
estimated this contribution to be lower than $\sim30\%$ of the total $\gamma$GDE. Here, as reference we use the estimated flux of unresolved sources derived in Ref.~\cite{Eckner_Limits}, which is shown summed to the truly diffuse emission (i.e. originated from CR interactions) to give the cyan line shown in the left panel.
{We notice, however, that the contribution from pulsar-powered ~\cite{Vecchiotti_2022} or  star forming regions and hypernovae ~\cite{Fang:2021ylv} to the diffuse flux, although yet very uncertain, could be even dominant. The fact that these contributions can be much larger could be in tension not only with the $\gamma$-optimized model, but also with the Base model.}  

\end{itemize}

The limited statistics does not allow to claim a preference for a class of model with respect to the other. Both the Base and the gamma-optimized scenarios -- both in the ``Min'' and ``Max'' versions -- appear compatible with the data. 
We can however safely conclude that truly diffuse emission must dominate the emission, with a subdominant (but non-negligible) contribution from unresolved sources.


In \cite{TibetASgamma:2021tpz} the measured spectra were compared with a conventional and a spatial dependent analytical CR models presented in Ref.\cite{Lipari:2018gzn}.
In the inner region (A) Tibet results looked in slightly better agreement with the spatial dependent scenario although the scatter of the data points and their large errors did not allow a conclusive preference of one models over the other.

A similar uncertain situation was found in Ref.\cite{Luque:2022buq,DelaTorreLuque:2022ats}, where the Base and $\gamma$-optimized Min and Max models -- which assume more realistic source spectra above 10 TeV respect to Ref.\cite{Lipari:2018gzn} -- were compared with Tibet data. Only the Max setups, however, were found to be compatible with Tibet results.


\subsection{Comparison with LHAASO results}
\label{sec:LHAASO_comp}

The Large High Altitude Air Shower Observatory
(LHAASO) is a large area, wide field-of-view observatory for CRs and $\gamma$-rays using a hybrid detection technique~\cite{LHAASO:2019qtb}.
This experiment combines measurements taken with the 78,000 m$^2$ wide Kilometer-2 Array (KM2A) charged particle detector with those of a Water Cherenkov Detector Array (WCDA) allowing to cover an energy interval going from few TeV up to PeV.  
While the field of view (2 sr) and energy and spatial resolutions are comparable to those of Tibet, its sensitivity allows LHAASO to collect much more events and have measurements with high significance in much shorter times. 

 

After releasing preliminary measurements~\cite{Zhao:2021dqj}, 
 the LHAASO collaboration presented their observations of the average diffuse flux in two regions covering the Galactic plane -- the inner ($15^{\circ} < l < 125^{\circ}$) and outer ($125^{\circ} < l < 235^{\circ}$) regions ($|b| < 5^{\circ}$) -- from $10$~TeV to $1$~PeV obtained with the KM2A array~\cite{LHAASO:2023gne}. 
More recently the collaboration updated those results including also the $\gamma$GDE flux measurement performed with the WCDA from $1$ to $25$~TeV, which led to the release of wide-band spectral distributions in the inner and outer regions from 1 TeV to 1 PeV~\cite{LHAASO:2024lnz}.
Interestingly, the WCDA spectra are consistent with a power-law with indices of $-2.67 \pm 0.05{~\rm stat}$ and $-2.83 \pm 0.19{~\rm stat}$ in the inner and outer regions, respectively, hinting (with a significance of $\sim2\sigma$) that the primary CR spectrum may get harder toward the GC.
However, we will see that current LHAASO data do not provide enough discriminating power to robustly confirm or constrain spatially-dependent CR propagation models at the moment.

The fluxes of the diffuse emission were found to be higher by a factor of $1.5 - 2.7$ than a simple reference model that assumes local values for the CR flux and gas density across the Galaxy. 
However, more realistic conventional models of the $\gamma$GDE (as, for instance, our Base models) that take into account the increase of the number of CR sources and gas densities towards the GC~\cite{Evoli_2007}, predict significantly higher $\gamma$GDE fluxes which are still compatible with LHAASO results (see Fig. 4 in Ref. \cite{DelaTorreLuque:2022ats}, and Fig.\ref{fig:Full_LHAASO_Comp} in the Appendix \ref{appendix:HE_comparisons}, as well as the detailed study of Ref.~\cite{Vecchiotti:2024kkz}).

The main reason making LHAASO data still insufficient to discern between inhomogeneous and uniform models of CR propagation is the  mask that they use to remove the emission from sources (as also noted in Ref.~\cite{Vecchiotti:2024kkz}). Their procedure requires masking a significant area around the sources listed in the TeVCat~\cite{2008ICRC....3.1341W} and those detected by LHAASO~\cite{LHAASO:2023rpg}.  
This results in quite a severe reduction of the portion of the GP observed (see Fig. 1 of Ref. \cite{LHAASO:2024lnz}), that especially cuts away most of the inner $2^{\circ}$ around the GP, where the GDE flux is predicted to be higher and where the difference with non-uniform CR propagation models is expected to be most prominent. 
As a consequence the $\gamma$GDE flux measured by LHAASO along the GP is lower than the real one (see {\rm e.g.} Fig.\ref{fig:LHAASO_WCDA_Inner} ) and a consistent comparison with the measurements requires applying the mask also to the predicted $\gamma$GDE.


\begin{figure}[t]
\centering
\includegraphics[width=0.495\textwidth, height = 0.22\textheight]{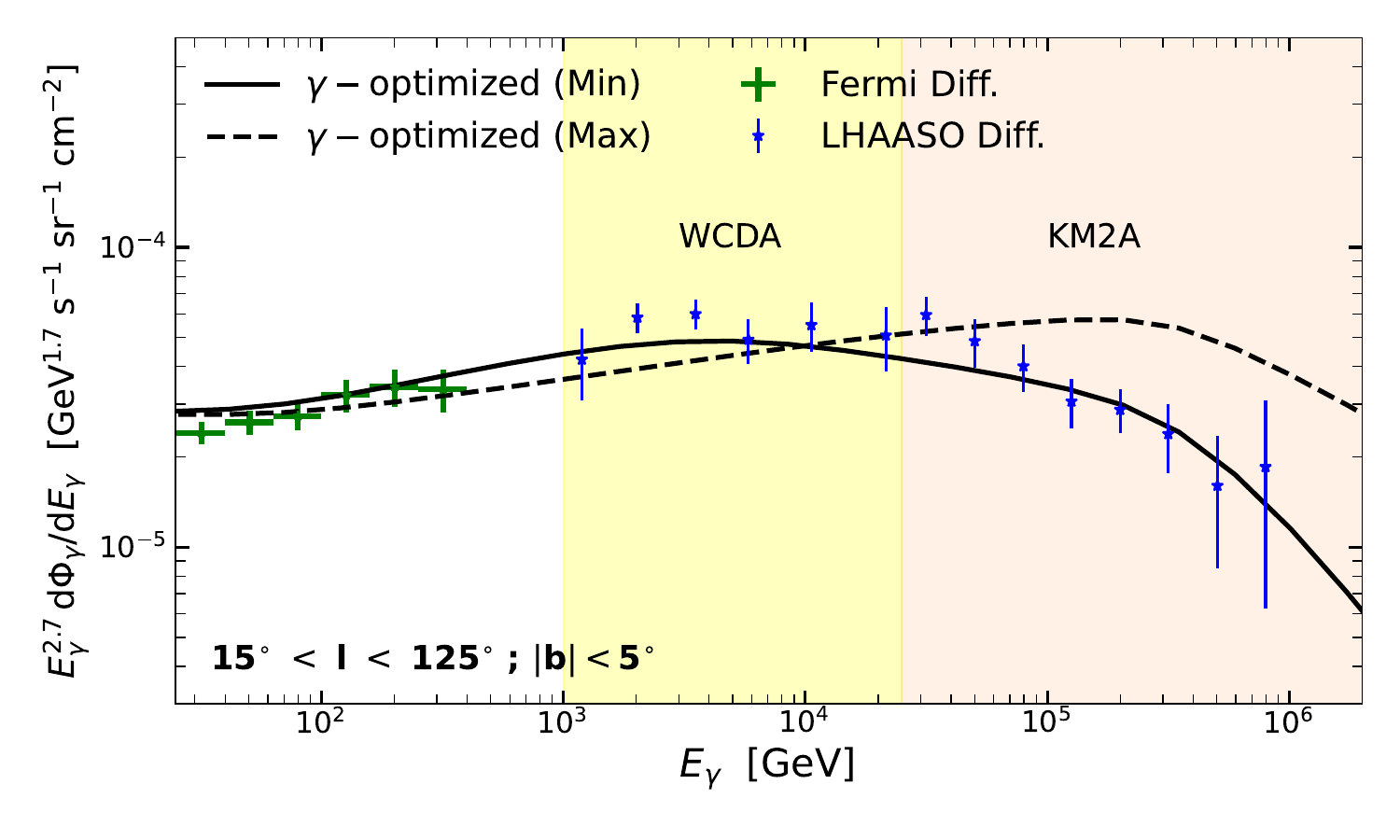}
\includegraphics[width=0.495\textwidth, height = 0.22\textheight]{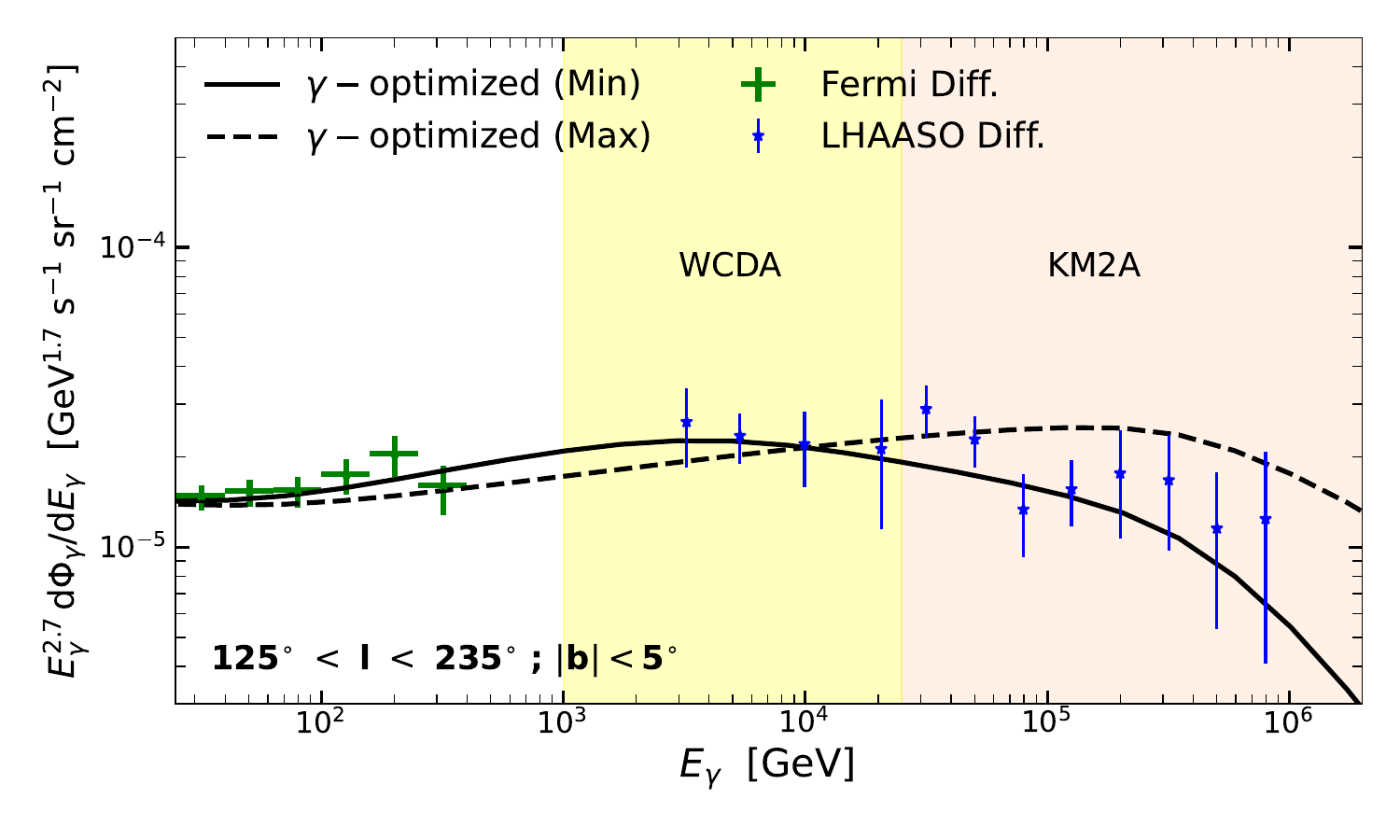}
\caption{Comparison of the LHAASO (KM2A  and newest WCDA) diffuse spectrum \cite{LHAASO:2024lnz} with the $\gamma$GDE computed with the $\gamma$-optimized Min and Max models (for the comparison with the corresponding Base models see Fig.\ref{fig:Full_LHAASO_Comp}). We shaded the energy region covered by the WCDA and KM2A in different colors, for clarity. We obtained Fermi-LAT diffuse data points applying the LHAASO mask and subtracting contribution from sources (4FGL catalog) and the EGB. Note as the models do not represent a fit of LHAASO data rather they are predicted on the basis of CR and lower energy $\gamma$-ray data. }
\label{fig:LHAASO_optimized_Comp}
\end{figure}

In Fig.\ref{fig:LHAASO_optimized_Comp} we plot the wide-band spectra of the $\gamma$GDE predicted with the $\gamma$-optimized Min and Max models, in comparison the Fermi-LAT and LHAASO WCDA+KM2A data in the inner (left panel) and outer (right panel) regions. The Fermi-LAT measurements shown here are those applying the LHAASO mask and subtracting contribution from sources in the 4FGL catalog and the EGB. As we see, the agreement between the Min model predictions and the data, over more than four energy decades, is quite remarkable bearing in mind that this is not the result of a fit but is a prediction based on CR and Fermi data.
In particular, the reduced $\chi^2$ value (computed as $\chi^2$ over number of LHAASO data points) is of $0.8$ for the Min model and 9.5 for the Max model, in the inner region, and of $0.6$ and $1.9$ for the Min and Max models, respectively, in the outer region. This means, more than $10\sigma$ evidence that the Max model is incompatible with LHAASO data. This result leads to the conclusion that the IceTop measurements of the local CR spectrum are incompatible with LHAASO-KM2A data if its shape is representative of the large scale Galactic CR population.
{We note that, in these predictions, the CR break at $\sim300$~GV is implemented as a break in the injection spectrum, while there are hints favoring that this break is due to an actual break in the diffusion coefficient. Therefore, we also estimated the expected difference in the diffuse flux when adopting such a break in the diffusion coefficient. From this exercise, we found differences in the predicted gamma-ray and neutrino fluxes are below $15\%$ (larger flux) at $100$~TeV, revealing that other sources of uncertainty, such as the fit of the primary CR spectrum or cross sections, are the dominant ones. However, the spatial behavior of such break is challenging to assess, and beyond the goals of this paper.}


\begin{figure}[!t]
\centering

\includegraphics[width=0.49\textwidth]{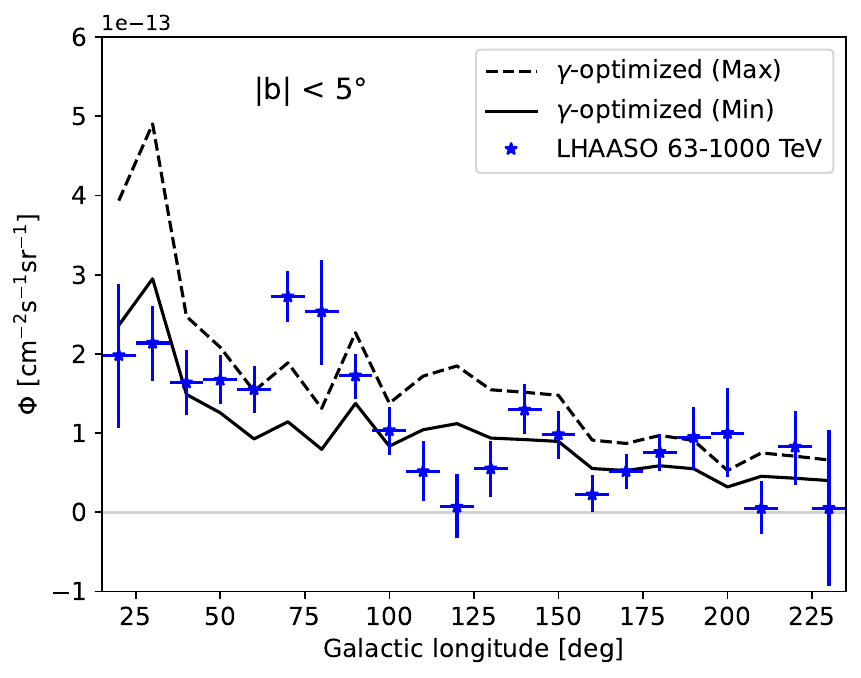}
\includegraphics[width=0.49\textwidth]{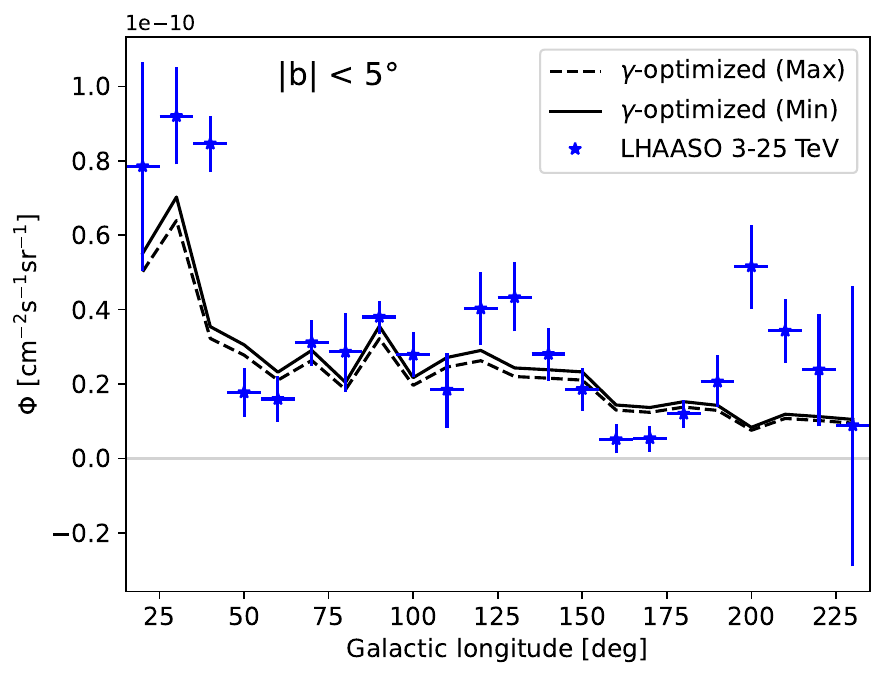}

\includegraphics[width=0.49\textwidth]{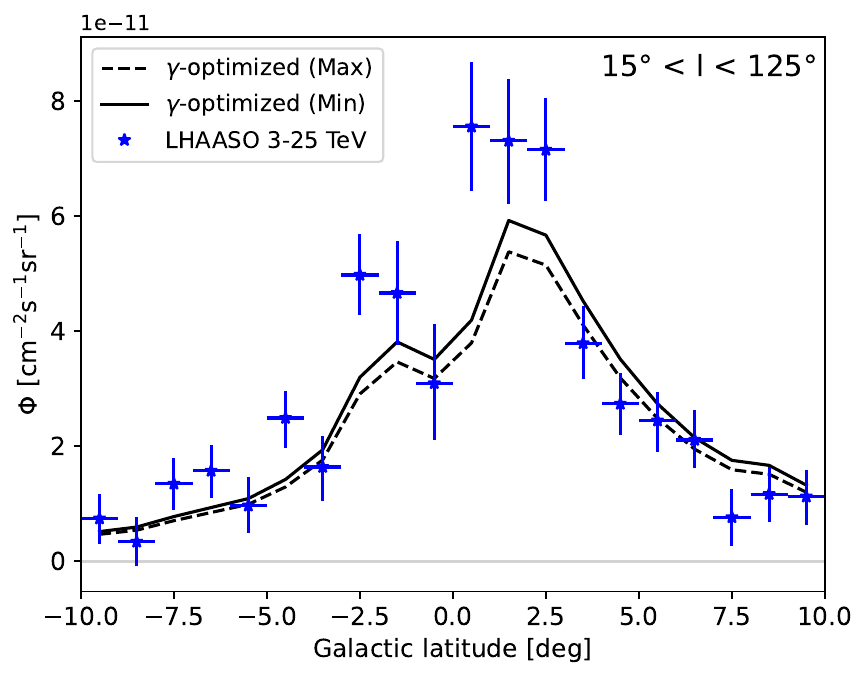}
\includegraphics[width=0.49\textwidth]{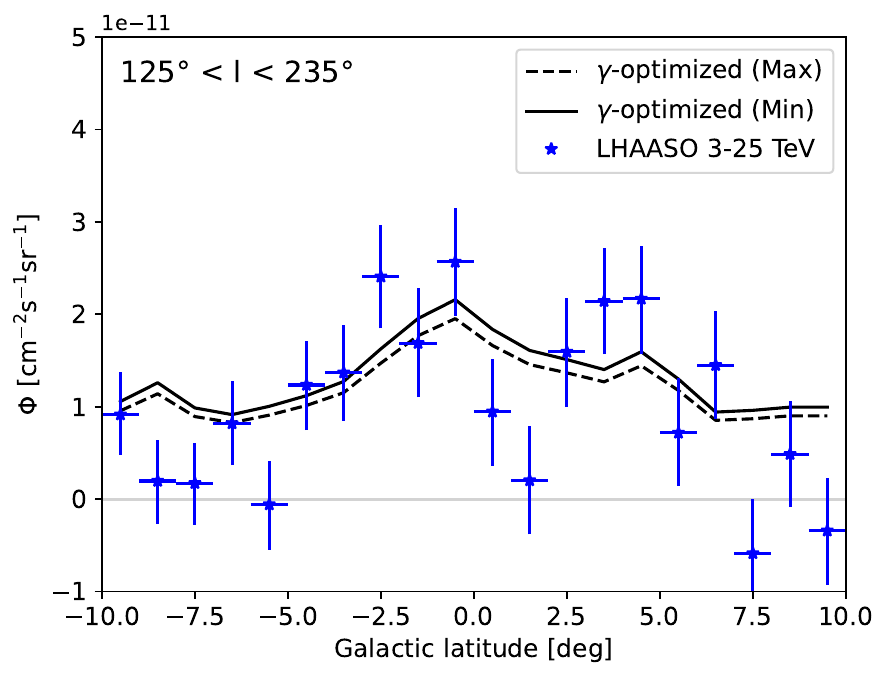}
\caption{LHAASO longitude (top panels) and latitude (bottom panels) profiles of the $\gamma$GDE, compared to the prediction from the $\gamma$-optimized models. The LHAASO mask is applied to the models in every case. 
}
\label{fig:Profs}
\end{figure}

Moreover, the contribution from unresolved sources (which are expected to be more concentrated at low latitudes) is heavily suppressed with respect to the truly diffuse emission --arising from the interactions of the CR sea -- as a consequence of the masking technique adopted by the LHAASO collaboration.
There have been a few recent papers estimating the contribution from unresolved sources to the LHAASO masked data~\cite{Vecchiotti:2024kkz, He:2025oys,Lipari:2024pzo}. They all agree with the fact that the contribution from sources must be lower than $\sim10-30\%$, depending on the energy range. In particular, Ref.~\cite{He:2025oys} claimed that a model with homogeneous diffusion could not explain the LHAASO measurements (namely, from the WCDA) in the inner Galaxy with the expected contribution of unresolved sources.
Taking the most constraining data point (exactly at $125.9$~TeV), we estimate that the maximum contribution from unresolved sources in the inner region is of $23\%$ at the $2\sigma$. However, at around $10$~TeV, this contribution can be higher than $40\%$ and be still compatible within $2\sigma$ with LHAASO data.
{In turn, other studies have proposed that the contribution from TeV halos (see e.g. Refs.~\cite{Dekker_2024} and~\cite{Yan:2023hpt} and references therein) could become very important at TeV energies, when assuming that all middle aged pulsars develop a Geminga-like TeV halo (which is in conflict with Ref.~\cite{Martin_2022}). Similarly, Ref.~\cite{Menchiari:2024uce} studied the potential contribution of massive stellar clusters to the $\gamma$-ray diffuse sky, finding that this contribution may be very important at TeV energies, but that it should become almost negligible at the PeV.}

In Fig.\ref{fig:Profs} we also compare the longitude and latitude emission profiles of the Min and Max models with LHAASO results in the energy bands $3-25$ TeV and $63-1000$ TeV. Again here, we find that the emission measured by KM2A in the innermost region clearly favor the Min setup (see top left panel of Fig.~\ref{fig:Profs}). 
We also remark here that the small underestimation of the data towards the GC in the energy range from $3$~TeV to $25$~TeV (top-right and bottom-left panels) may indicate that the contribution from unresolved sources grows as approaching the GC, which is actually expected (see, e.g. Refs.\cite{Lipari:2024pzo, Kaci:2024lwx}).
Some papers have also considered the contribution from TeV halos in the diffuse TeV $\gamma$-ray sky~\cite{Dekker_2024, Yan:2023hpt}. However, we note that most of these estimations assume that all middle aged pulsars in the ATNF catalog develop a TeV halo with parameters similar to describing the Geminga TeV halo. Since only a few TeV halos (only 3 clear ones)~\cite{Amato:2024dss} have been detected, we expect this estimation to represent an overestimation, since it is possible that not all middle-aged pulsars result in TeV halos and probably taking the Geminga parameters is also an optimistic assumption (see Ref.~\cite{Martin_2022}). 

Turning our discussion now to the discrimination between the $\gamma$-optimized and Base models (i.e. inhomogeneous vs uniform CR propagation, respectively) with LHAASO data, we find (as already found in Ref.~\cite{Vecchiotti:2024kkz}) that, due to the mask, the differences between both scenarios become too low to make a clear statement only from the inner and outer regions (see Fig.~\ref{fig:Full_LHAASO_Comp}). Interestingly, these measurements seem to hint to a different spectral index for the the power-laws describing the detected flux in each region, but with negligible significance yet.  
Potentially, having measurements focused on more internal regions could allow for a discrimination between both scenarios. 
Thanks to the large statistics collected with the WCDA, besides the inner and other regions, LHAASO was also able to measure the spectra of the $\gamma$GDE in three sub-regions of the inner region with $15^{\circ} < l < 50^{\circ}$, $50^{\circ} < l < 90^{\circ}$ and $90^{\circ} < l < 125^{\circ}$. Again here, the LHAASO collaboration points to a mild, $\sim 2 \sigma$ hint favoring a spectral index varying with the Galactic position.
The two innermost among these sub-regions have the largest statistics and are most sensitive to the choice of the CR propagation setup. 
In the left panel of Fig.~\ref{fig:LHAASO_WCDA_Inner} we compare the $\gamma$-optimized and Base models predictions, both in the Min setup, with LHAASO-WCDA results in the most internal of those regions. Although here we appreciate a larger difference between $\gamma$-optimized and Base models, their difference is still low, and the contribution from unresolved sources is difficult to be estimated.
In addition, these measurements still lack systematic uncertainties. 
We also note that both scenarios follow very well the trend of the data, possibly pointing again to a subdominant contribution from unresolved sources.

In the left panel of this figure, we report the model predictions, compared to LHAASO and Fermi data, both obtained using the LHAASO mask. In the right panel, no mask is applied to the models and the Fermi data. It is evident that the LHAASO mask has quite an important impact on Fermi data, which significantly deviates from the truly diffuse model above $\sim100$~GeV. 


\begin{figure}[t]
\includegraphics[width=0.49\textwidth]{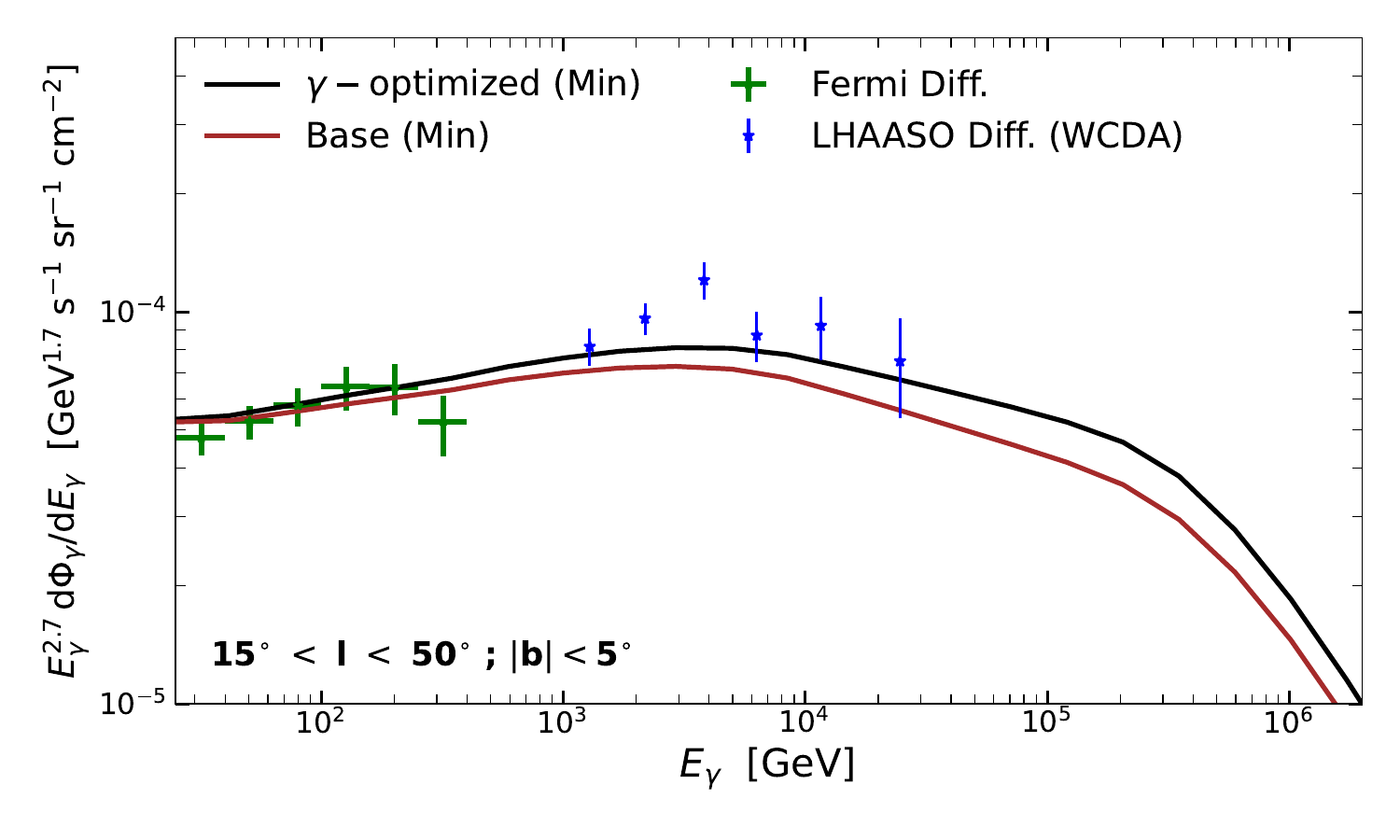}
\includegraphics[width=0.49\textwidth]{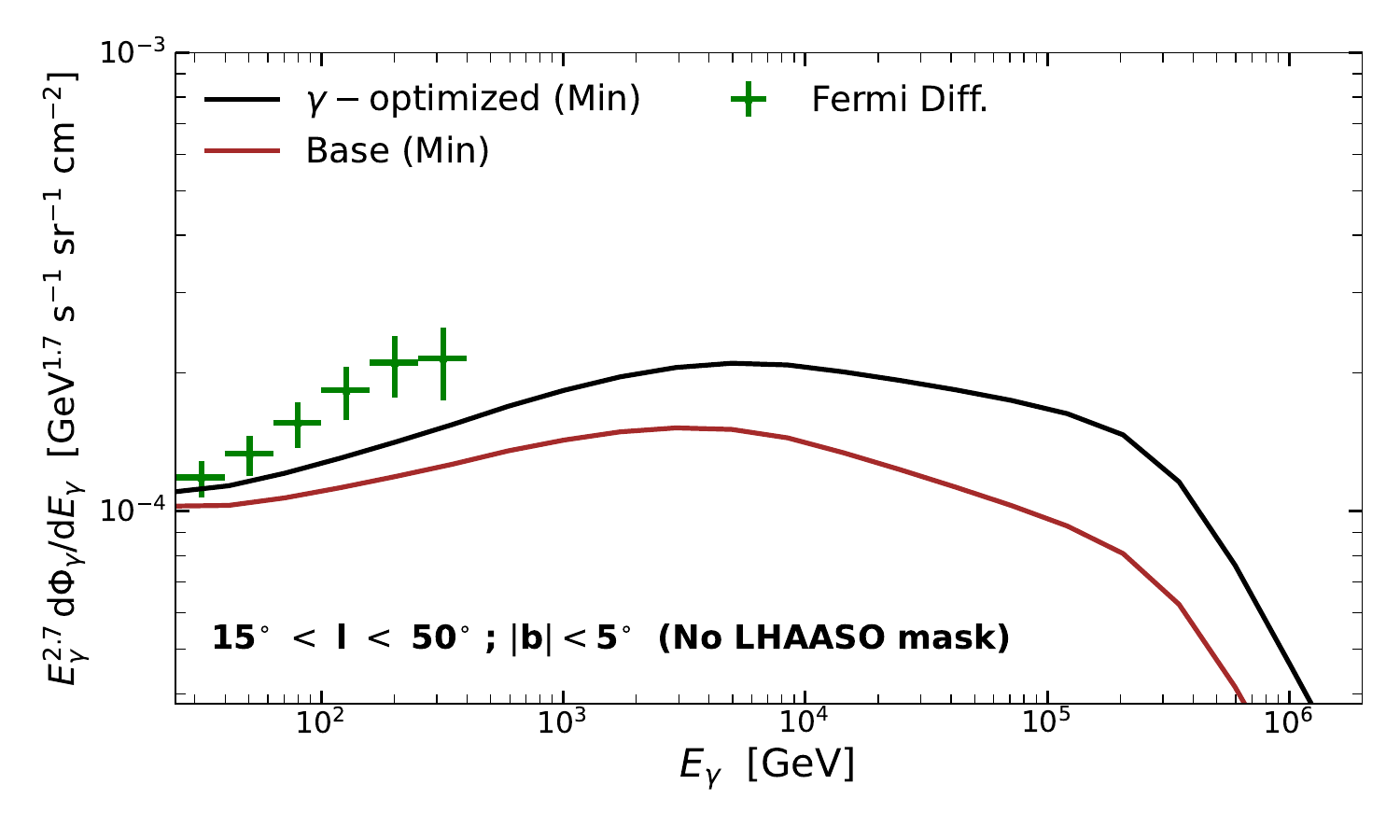}
\caption{Predicted spectra from the $\gamma$-\textit{optimized} and Base models in the Min configuration are compared to the recent diffuse LHAASO-WCDA data (Only statistical errors are available) and Fermi-LAT data (with errorbars indicating statistical and systematic uncertainties from the effective area) in the $15^{\circ} < l < 50^{\circ}$, $\vert b \vert < 5^{\circ}$ region. In the left panel the models and Fermi data are computed in the LHAASO ROI (masked). In the right panel no mask is used. }
\label{fig:LHAASO_WCDA_Inner}
\end{figure}

\section{Galactic neutrino emission from cosmic-ray interactions}
\label{sec:neutrinos}


As mentioned before, the template fitting analyses conducted during the last years by ANTARES and IceCube collaborations constrain different diffuse Galactic scenarios and highlight the presence of a signal excess when observing along the whole Galactic plane through the cascade events sample~\cite{Aguilar_2011, Visser_2015, IceCube:2023ame}.
Following the spatial and energy distribution of theoretical templates the analysis leaves as a free parameter the normalization of the energy spectrum. The observed neutrino excess results in different best-fit values of the normalization of the different templates.

\begin{figure}[t]
\includegraphics[width=0.495\textwidth]{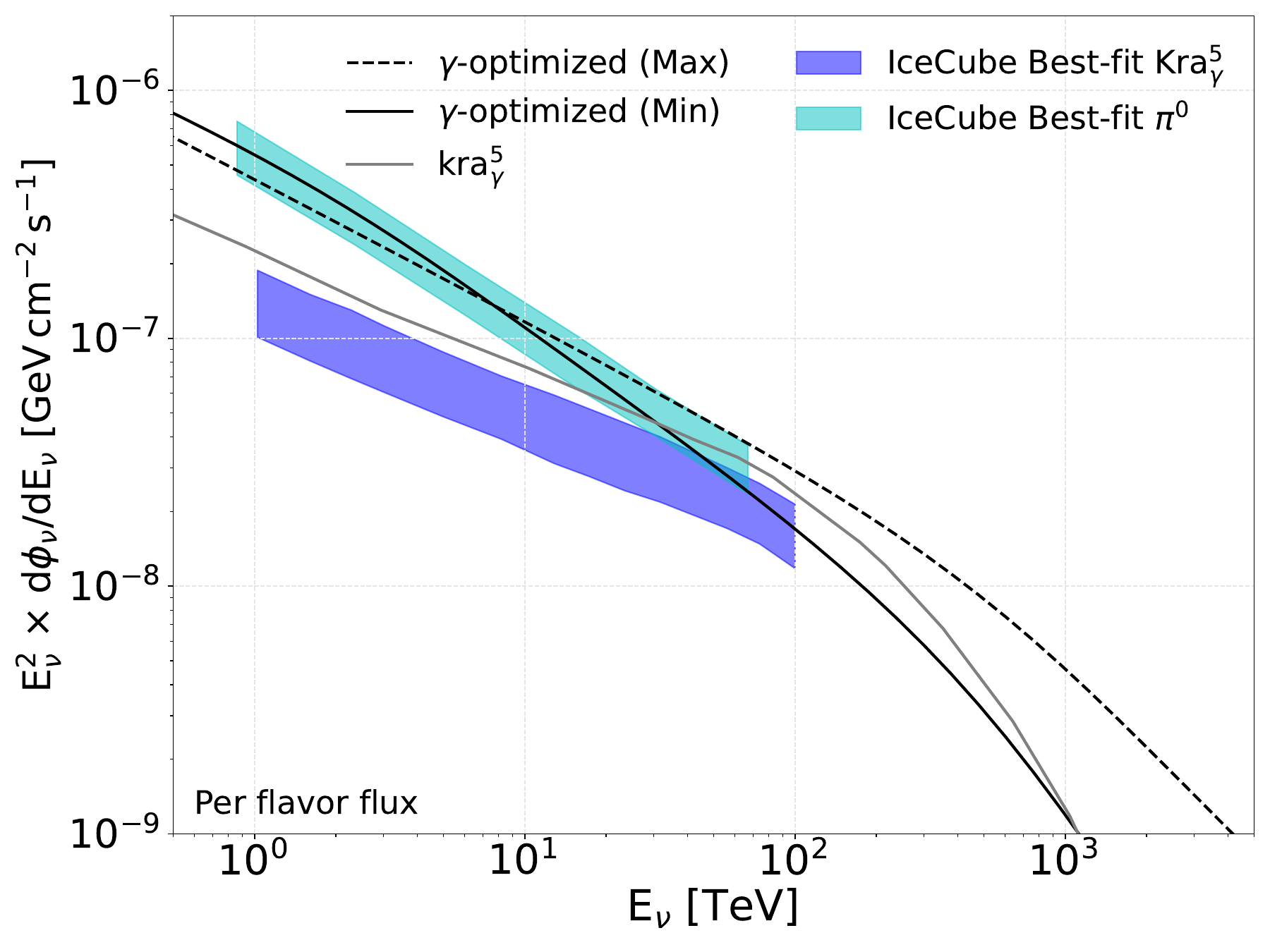} 
\includegraphics[width=0.495\textwidth]{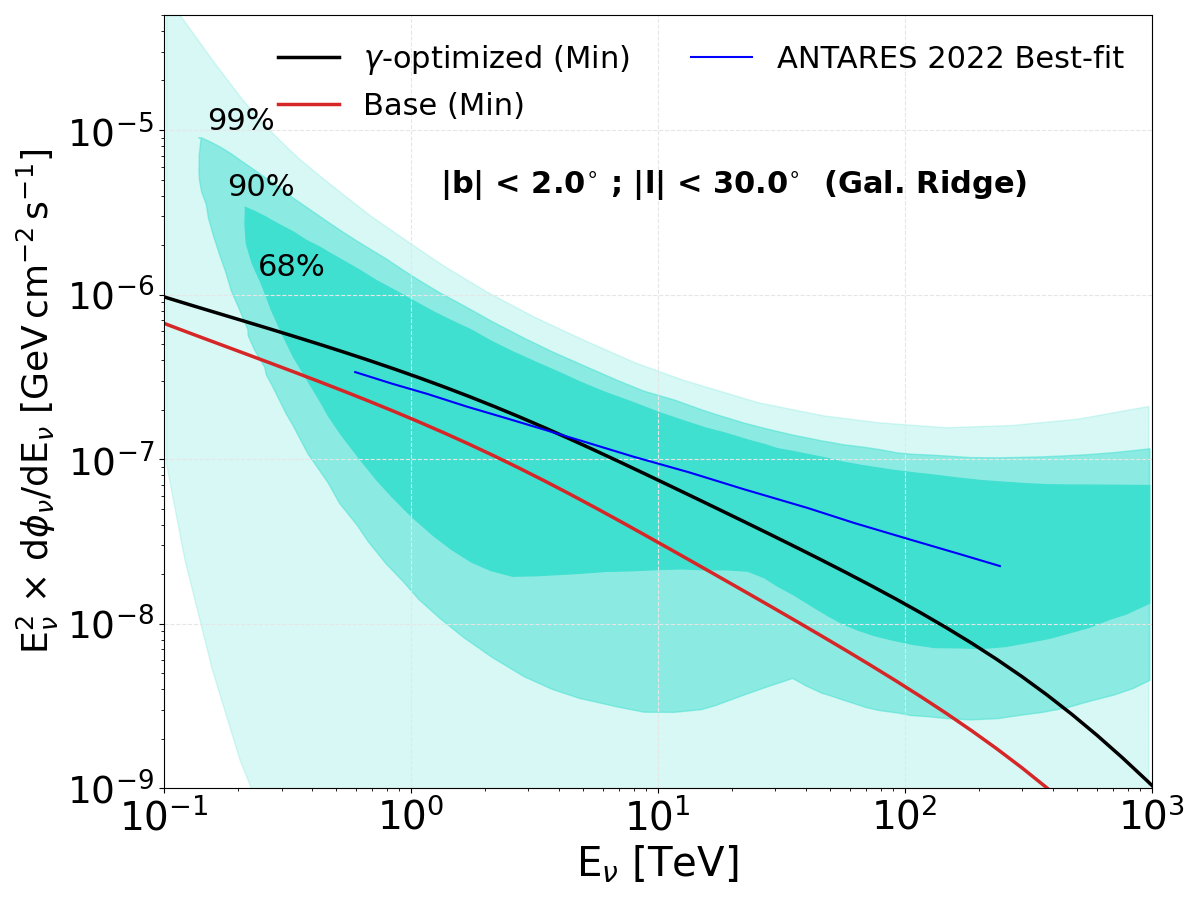}
\caption{\textbf{Left panel:} Predicted full-sky $\nu$ diffuse emission (per flavor) from the $\gamma$-\textit{optimized} model compared to the best-fit IceCube flux extracted from the KRA-$\gamma$ (cutoff energy of E$_c=5$~PeV) and $\pi^0$ models. The predicted flux from the KRA-$\gamma^5$ model is also reported as a gre line, for comparison. \textbf{Right panel:} Comparison of the predicted $\nu$-GDE from the $\gamma$-optimized (Min and Max) models with the per-flavour flux measured at the Galactic ridge ($|l|<30^{\circ}$ and $|b|<2^{\circ}$) by ANTARES~\cite{Albert_2023}.}  
\label{fig:IceCube}
\end{figure}

The most significant excesses have been obtained following the KRA$_\gamma^5$ and $\pi_0$ templates, respectively, with $4.37 \sigma$ and $4.71 \sigma$ of significance. 
The best fits for the KRA$_\gamma$ and $\pi_0$ templates are reported in the plot on the left of Fig.~\ref{fig:IceCube} with the blue and green bands. 
The $\pi_0$ best-fit model has the highest significance with respect to IceCube neutrino data. 
On the other hand, the IceCube~\cite{IceCube:2023ame} result shows that this neutrino flux normalization is a factor $\sim 4$ above the $\pi_0$ model expectations, making it markedly incompatible with Fermi-LAT data.
Moreover, the IceCube model independent pre-trial significance maps of the all-sky search and the best-fit spectral index angular distribution indicate that the emission is more concentrated and harder than predicted by that model (see the Supplementary material in~\cite{IceCube:2023ame}).

{
The KRA$_\gamma^5$ model is much closer to the best-fit IceCube observations.
Also for this template model the normalization of the IceCube best-fit is different, by a factor $0.55$ (hence less critically than the $\pi_0$ model), from that of the original model based on Fermi-LAT data.
We remind the reader that the KRA$_\gamma^5$ assumes a simplified exponential cutoff of the proton source spectrum at 5 PeV. 
A more realistic source spectrum may reach even better agreement with IceCube results, as implemented in the $\gamma$-optimized Min model which, adopts a broken power-law source spectrum so to match the proton spectrum measured by CALET~\cite{CALET:2019bmh}, DAMPE~\cite{DAMPE:2019gys} as well as KASCADE~\cite{Apel:2013uni} from 10 TeV to 10 PeV at least.
In fact the $\gamma$-optimized Min predicts a smaller neutrino flux above 10 TeV -- where the signal overcome the atmospheric $\nu$ flux hence giving the largest contribution to the detection significance -- which is key in the analysis performed by the IceCube collaboration.
}


In the left panel of Fig.~\ref{fig:IceCube} we compare our models with the best-fits quoted by IceCube. Considering that the reported fluxes are full-sky integrated, this comparison does not account for the spatial distribution of expected neutrino flux.
Interestingly, both Min and Max setups are consistent with the $\pi_0$ IceCube best-fit, that with the highest significance.
{The Min, however, should be preferred since it is in better agreement with GDE spectrum measured by LHAASO.}
For this reasons, we suggest adopting the updated $\gamma$-optimized Min template in future template fitting analyses.



Here, we also extend the work done in Ref.~\cite{DelaTorreLuque:2022ats} comparing the expected neutrino SEDs of $\gamma$-optimized models with the recent results obtained by the ANTARES collaboration looking at the so called Galactic Centre ridge ($|l|<30^{\circ}$ and $|b|<2^{\circ}$)~\cite{Albert_2023}. In that work an on/off analysis for this central part of the Galaxy was performed. The resulted best fit (under a power-law SED assumption) is here reported in the right plot of Fig.~\ref{fig:IceCube}. As shown in that figure, the $\gamma$-optimized models are in good agreement with the ANTARES best-fit emission and inside the $1\sigma$ reported contour from 500 GeV to 500 TeV. While the uncertainties in this detection are still too high, our results can be considered an important independent cross-check of the consistency of our models with neutrino observations.
In the left panel of Fig.~\ref{fig:IceCube_App}, we show a similar comparison as in the left panel of Fig.~\ref{fig:IceCube} using a different scale.
In the right panel of Fig.~\ref{fig:IceCube}, we show the difference between the predicted emission in the Galactic Ridge from the $\gamma$-optimized and Base models (Min, in both cases). Here, the difference between both scenarios is much more significant, and future data could be able to distinguish between them. See also the left panel of Fig.~\ref{fig:IceCube_App}, where we report a comparison between the Base and $\gamma$-optimized models for the IceCube full-sky analysis.

In addition, future observations of the KM3NeT telescope will have the capability to resolve
peculiar regions of the GP -- like the Central Molecular Zone -- where the most massive molecular clouds are placed. These future observations can discriminate among different diffuse scenarios and highlight possible PeVatron-related emission components. 
However, we remark that the comparison of the $\gamma$-optimized models with neutrino data analyses already shows that a potential local contribution of Galactic PeVatrons is expected to be subdominant with respect to the diffuse Galactic sea emission (when a large enough portion of the Galaxy is considered).

\newpage

\section{Discussion and conclusions}
\label{sec:discussion}
In this paper we consider a set of models for the CR distribution in the Galaxy and the associated high-energy emission in the $\gamma$-ray and neutrino channels. These numerical models were presented first in Refs.~\cite{Luque:2022buq, DelaTorreLuque:2022ats}, and are computed numerically adopting the {\tt DRAGON} and {\tt HERMES} codes. The approach we follow is to tune the models on the locally measured CR data (bracketing the uncertainty in the high-energy domain by considering a Min and Max configurations), and consider different transport setups: a ``conventional'' one (Base) with homogeneous diffusion, and a $\gamma$-optimized scenario characterized by a harder spectrum in the inner Galaxy (as suggested by Fermi-LAT data).

This approach, with no further ``ad hoc'' tuning, provides a remarkable agreement with a wide set of data from different experiments ranging from GeV to PeV.
Our analysis confirms that the  $\gamma$GDE and $\nu$GDE are mostly originated by the so-called ``diffuse sea'' of Galactic CRs all the way up to the largest detected energies.
The agreement between the models and data is not trivial at all --- especially at the high energies we consider here ---  given the potential role of source stochasticity~\cite{Stall_2023} and the debate about the scarcity and short duration of Galactic PeVatrons\cite{Kaci_2025}.
This success confirms the viability of $\gamma$-ray astronomy to probe the properties of the diffuse sea of CRs across the whole Galactic plane.

We find that the spectral steepening of the LHAASO data at $\simeq 100$ TeV (see Fig.~\ref{fig:LHAASO_optimized_Comp}) is well reproduced by the Min setup, which was tuned to match the local CR proton and Helium spectra at $\simeq 1$ PeV/n as observed by KASCADE \cite{Apel:2012tda,Apel:2013uni} , as well as by DAMPE~\cite{DAMPE:2019gys} and CALET~\cite{CALET:2019bmh} at lower energy. This finding supports the interpretation of the locally observed ``knee'' in the CR spectrum as a feature that characterizes the whole Galactic CR sea. A similar consideration holds for the steepening observed at a few TeV -- more evident in the outer region -- which can be associated to the feature visible in the DAMPE and CREAM CR data (see Fig. \ref{fig:CRnuclei}) slightly above $10$ TeV/n.
Conversely, the Max setup (tuned on the IceCube~\cite{IceCube:2019hmk} CR dataset) is ruled out by LHAASO $\gamma$-ray data by more than 10 $\sigma$ when combining inner and outer Galaxy regions.

This conclusion is almost independent on the choice of the propagation model (Base or $\gamma$-optimized) since the LHAASO data cannot provide enough discriminating power to prefer a transport scenario over the other (see also Ref.~\cite{Vecchiotti:2024kkz}). 
{On the other hand, it should be noted that this finding is apparently at odds with the CR proton spectrum recently measured by LHAASO itself \cite{LHAASO:2025byy} which is higher and harder than KASCADE's and closer to the IceCube's one. As discussed in Ref.~\cite{Castro:2025wgf}, 
if the LHAASO's would be confirmed to be the correct CR proton spectra at the Earth, the discrepancy with the GDE measured by the same experiment could hardly be explained by other systematics and would possibly require to assume a spatial variation of the CR spectral shape. }

We notice that the LHAASO dataset features a larger normalization (by a factor 1.5$-$2.7) compared to the predictions of toy models that naively extend the local CR spectra all over the Galaxy. 
However, both our realistic Base scenarios (that correctly takes into account the inhomogeneous source distribution) and our spatially-dependent ``$\gamma$-optimized'' models are compatible with the data. The reason is that the LHAASO analysis masks the regions that would be more suitable to discriminate between these scenarios, i.e. the low-longitude regions closer to the Galactic plane (see Fig.\ref{fig:LHAASO_WCDA_Inner}).  
We notice, however, that LHAASO WCDA found anyhow a mild hint of a spatially-dependent GDE spectral index being $-2.67 \pm 0.05{~\rm stat}$ in the inner region ($15^{\circ} < l < 125^{\circ}$) and $-2.83 \pm 0.19{~\rm stat}$ in the outer one ($125^{\circ} < l < 235^{\circ}$).
Regarding the compatibility of LHAASO and Tibet-AS$\gamma$ data, even accounting for the effect of the different masking technique and different regions of interest, the Tibet data still seem to exceed LHAASO observations. Unresolved sources can potentially explain this discrepancy~\cite{Fang:2023ffx, Vecchiotti_2022}, although its significance is hard to assess with the information at our disposal. 

In the future, CTA~\cite{CTAConsortium:2023tdz, CTA} data will access to the region of energy where the discrepancy between the base and $\gamma$-optimized model is larger, with the necessary angular resolution. Therefore, we expect that CTA may confirm or constrain the scenario of a inhomogeneous diffusion where CR confinement is larger at higher energies and close to the Galactic Center.

Another upcoming experiment that can be extremely useful for this purpose is SWGO \cite{Albert:2019afb}, a planned next-generation wide-field $\gamma$-ray observatory to be built in the Southern Hemisphere. Thanks to the wide field of view and very high duty cycle, it will be even more suitable for the study of truly diffuse emission along the Galactic plane extending the energy range currently covered by Fermi-LAT.

Before these experiments will operate, the Astri Mini-Array~\cite{Vercellone_2022} observatory, operating in the range from $\sim1$ to $\sim200$~TeV, can also provide important details of the high-energy $\gamma$GDE and the search for PeVatrons, with appropriate spatial and energy resolution, and covering a wide field of view.

We notice that HAWC and Tibet observations require to add a subdominant  -- but non negligible -- component of unresolved sources, as illustrated in Figs.~\ref{fig:HAWC_Comp} and~\ref{fig:TIBET_Comp}.
{Several studies have been performed to investigate the contribution of several classes of sources to Tibet and LHAASO GDE \cite{Shao:2023aoi,Yan:2023hpt,He:2025oys,Menchiari:2024uce}} which support that conclusion
(see also \cite{Fang:2023ffx, Vecchiotti:2024kkz, Eckner_Limits}). 

In particular, the profiles measured by the WCDA detector (Fig.~\ref{fig:Profs}) suggest that in regions
close to the GC the truly diffuse emission (i.e. diffuse emission from CR interactions with ISM
gas) needs an additional contribution. 
%
We suggest that our models can be used as a background template in order to characterize the very-high-energy  source population (both point-like and extended) in the dataset provided by these experiments.

Turning our attention to the multi-messenger context,  we remark that neutrino data provide very useful complementary information.
{
First of all we remind to the reader that 
IceCube found already hints (although not statistically significant yet) of a hardening of the $\nu$-GDE in the innermost region of the GP in its model independent pre-trial significance maps of the all-sky search and the best-fit spectral index angular distribution (see the Supplementary material in~\cite{IceCube:2023ame}).}
Here we argue that the adoption of the $\gamma$-optimized Min model in future template-fitting analyses could eventually provide statistically significant evidence in favor of that scenario.
In particular, a crucial test to validate this model would be a template-fitting analysis that combines the IceCube cascade samples and the KM3NeT and ANTARES track-like samples starting from a minimal energy of $\simeq 500$ GeV.

As a final summary, we briefly recap the main findings of the present work:

\begin{itemize}
    \item We presented a class of models of the Galactic truly diffuse gamma-ray emission tuned on local CR data that are compatible with a wide range of observations from GeV to TeV.
    \item Fermi-LAT data suggest a transport setup based on a radial dependent diffusion coefficient featuring a harder spectrum in the inner Galaxy. The very-high-energy LHAASO data cannot distinguish between the $\gamma$-optimized and the conventional homogeneous models. 
    \item On the other hand, LHAASO $\gamma$-ray data allow to identify a strong preference in favour of the Min setup of both the conventional and the $\gamma$-optimized models, tuned on the local CR datasets provided by CALET, DAMPE and KASCADE (in the ``knee'' region). 
    \item Unresolved sources seem to play a relevant, but still sub-dominant, role in the interpretation of these observations.
    \item Current IceCube and ANTARES observations are compatible with a scenario in which the Galactic diffuse emission is dominated by the CR sea interactions, with a slightly preference for the KRA$_\gamma$ scenarios when expectations and flux best-fit normalizations are compared.  Measurements towards the Galactic Centre are needed: KM3NeT will provide a very valuable data sample for future template analyses since 2/3 of the Galactic plane will be covered through track-like events which allow a better angular resolution. 
\end{itemize}

\newpage
\acknowledgments
We thanks Carmelo Evoli, Paolo Lipari, Giulia Pagliaroli and Silvia Vernetto, Sei Kato and Qiang Yuan for enlightening discussions.
P.D.L. is supported by the Juan de la Cierva JDC2022-048916-I grant, funded by MCIU/AEI/ 10.13039/501100011033 European Union "NextGenerationEU"/PRTR. The work of P.D.L. is also supported by the grants PID2021-125331NB-I00 and CEX2020-001007-S, both funded by MCIN/AEI/10.13039/501100011033 and by ``ERDF A way of making Europe''. P.D.L. also acknowledges the MultiDark Network, ref. RED2022-134411-T. This project used computing resources from the Swedish National Infrastructure for Computing (SNIC) under project No.2022/3-27 partially funded by the Swedish Research Council through grant no. 2018-05973. 
D.Gaggero and D. Grasso acknowledge support from the project ``Theoretical Astroparticle Physics (TAsP)'' funded by INFN.
D.~Grasso is seconded to the Embassy of Italy in the Kingdom of The Netherlands, The Hague.

\appendix

\section{Cosmic-ray local fits}
\label{appendix:CR}

Here we report a comparison of the CR proton and helium local spectra computed with our $\gamma$-optimized and Base models (coincident at our local position) in the Min and Max setup with a wide set of updated experimental data (see citations in Sec.~\ref{sec:CR_nuclei}). 

\begin{figure}[h]
\includegraphics[width=0.49\textwidth]{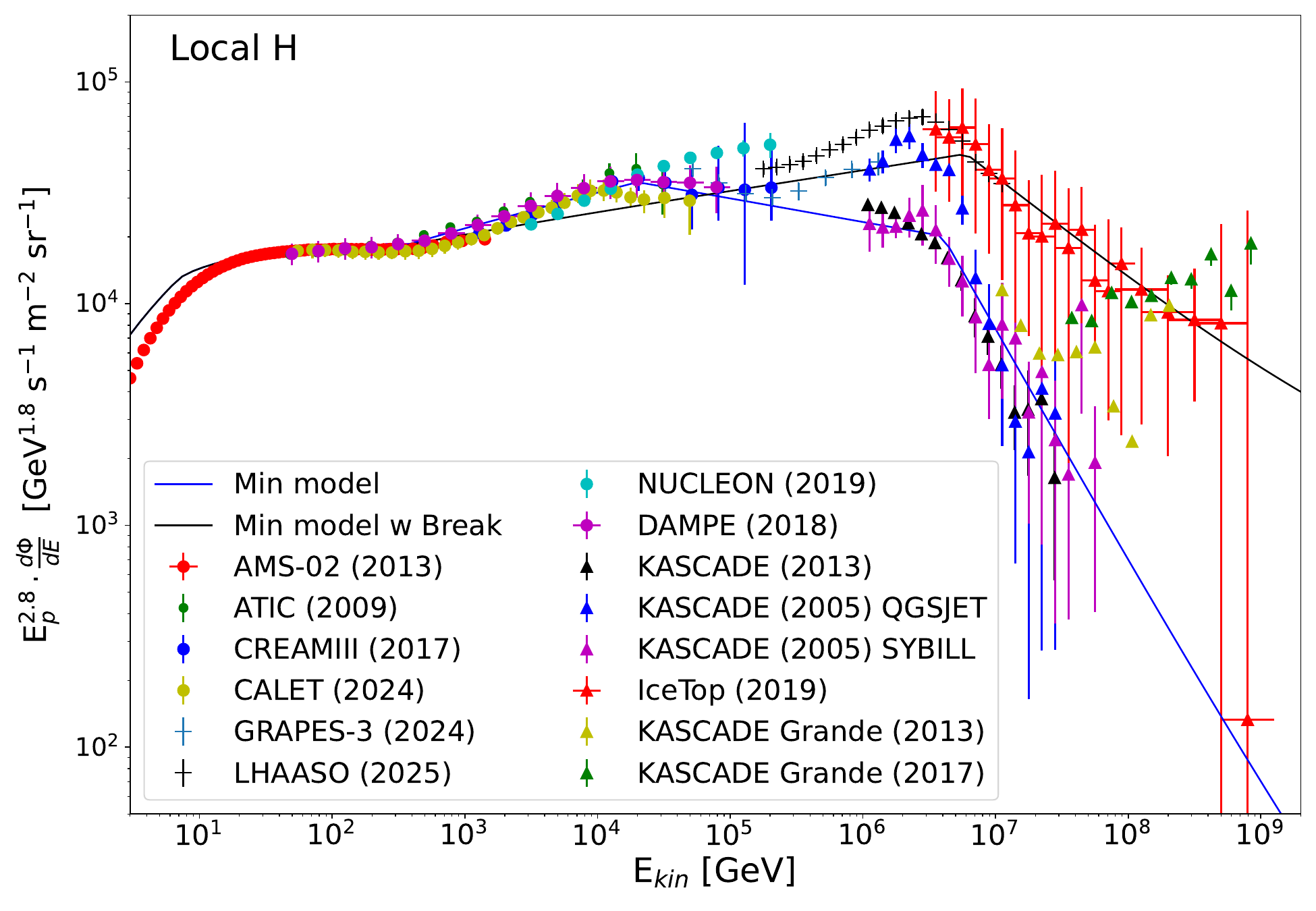}
\includegraphics[width=0.49\textwidth]{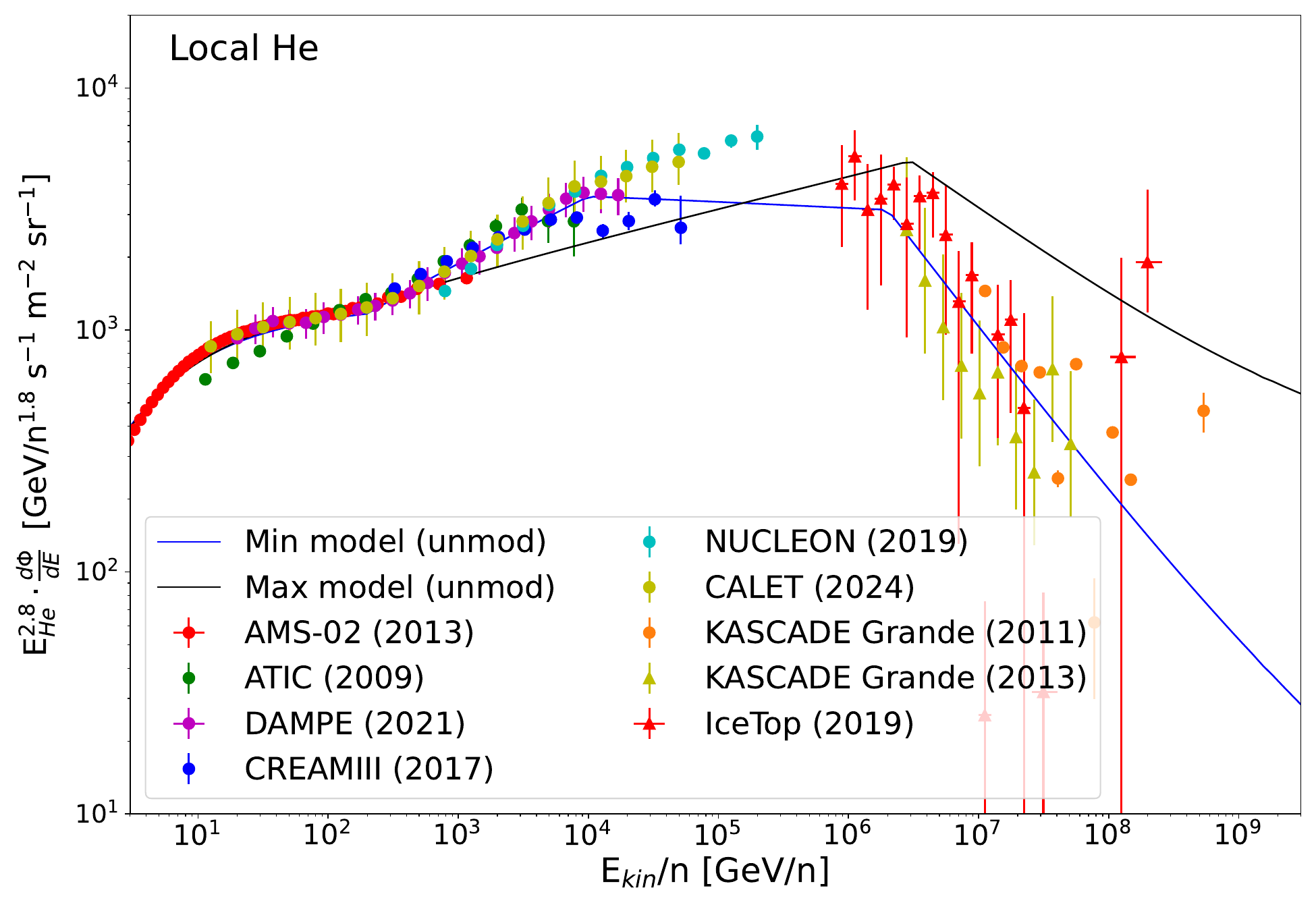} 
\caption{\textbf{Left panel:} CR proton spectral data against $\gamma$-optimized and Base (coincident) Min e Max models at our local position. \textbf{Right panel:} Same but for CR helium. 
}
\label{fig:CRnuclei}
\end{figure}

\newpage

\section{Components of the diffuse emission - Comparison with Fermi-LAT data}
\label{appendix:Fermi_Contrib}

\begin{figure}[!ht]
\centering
\includegraphics[width=0.48\textwidth]{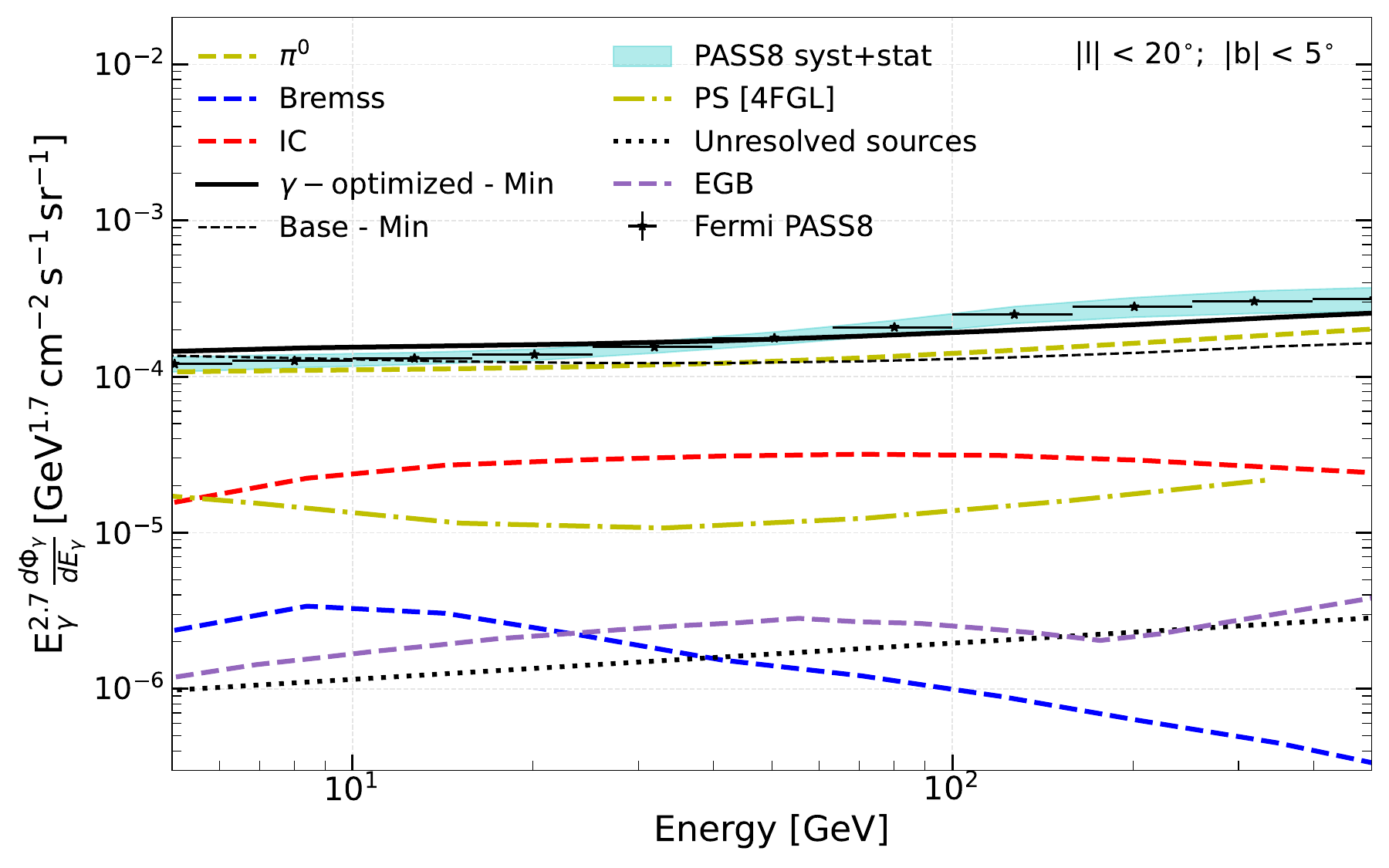} 
\hspace{0.1cm}
\includegraphics[width=0.48\textwidth]{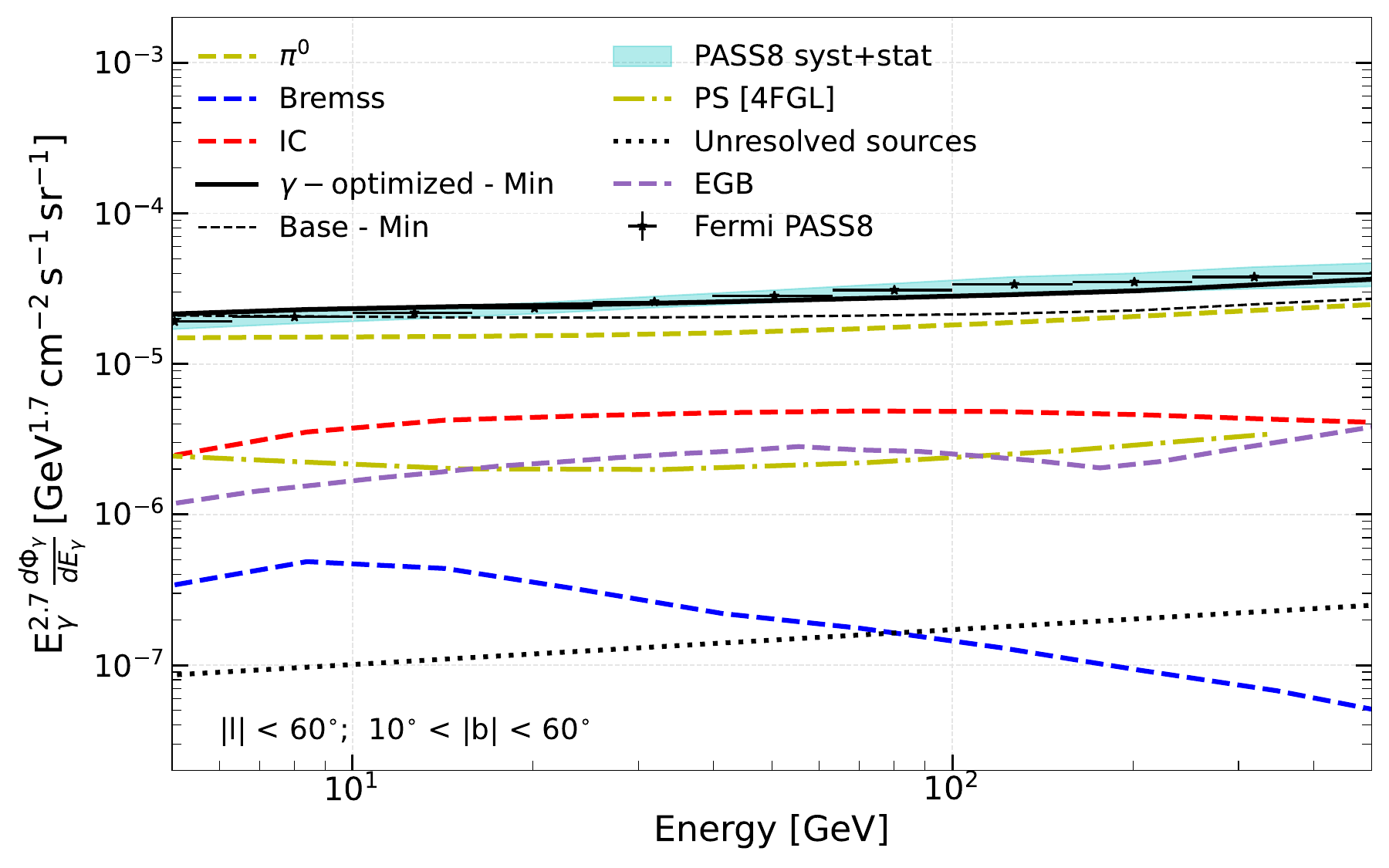}
\includegraphics[width=0.48\textwidth]{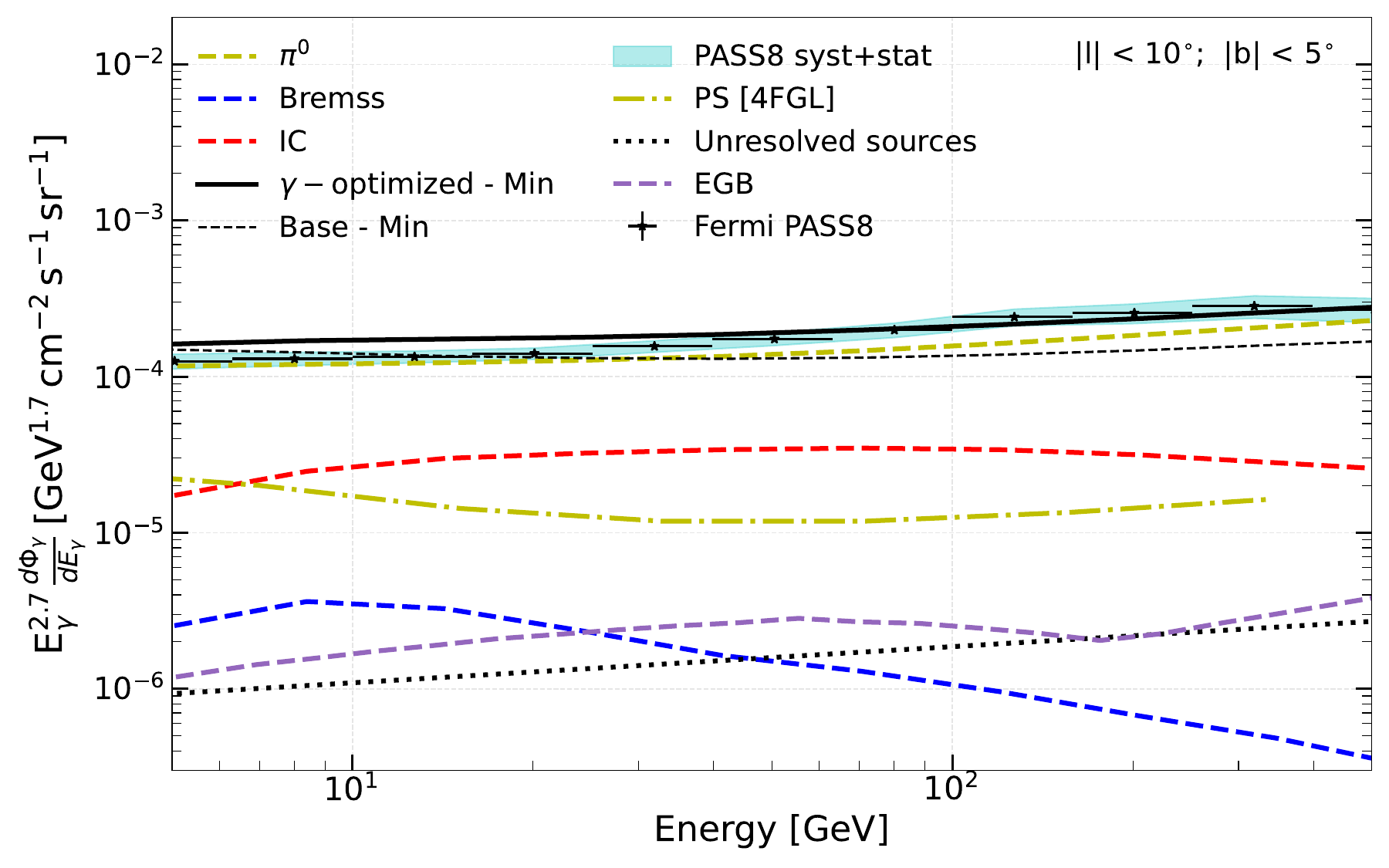}
\hspace{0.1cm}
\includegraphics[width=0.48\textwidth]{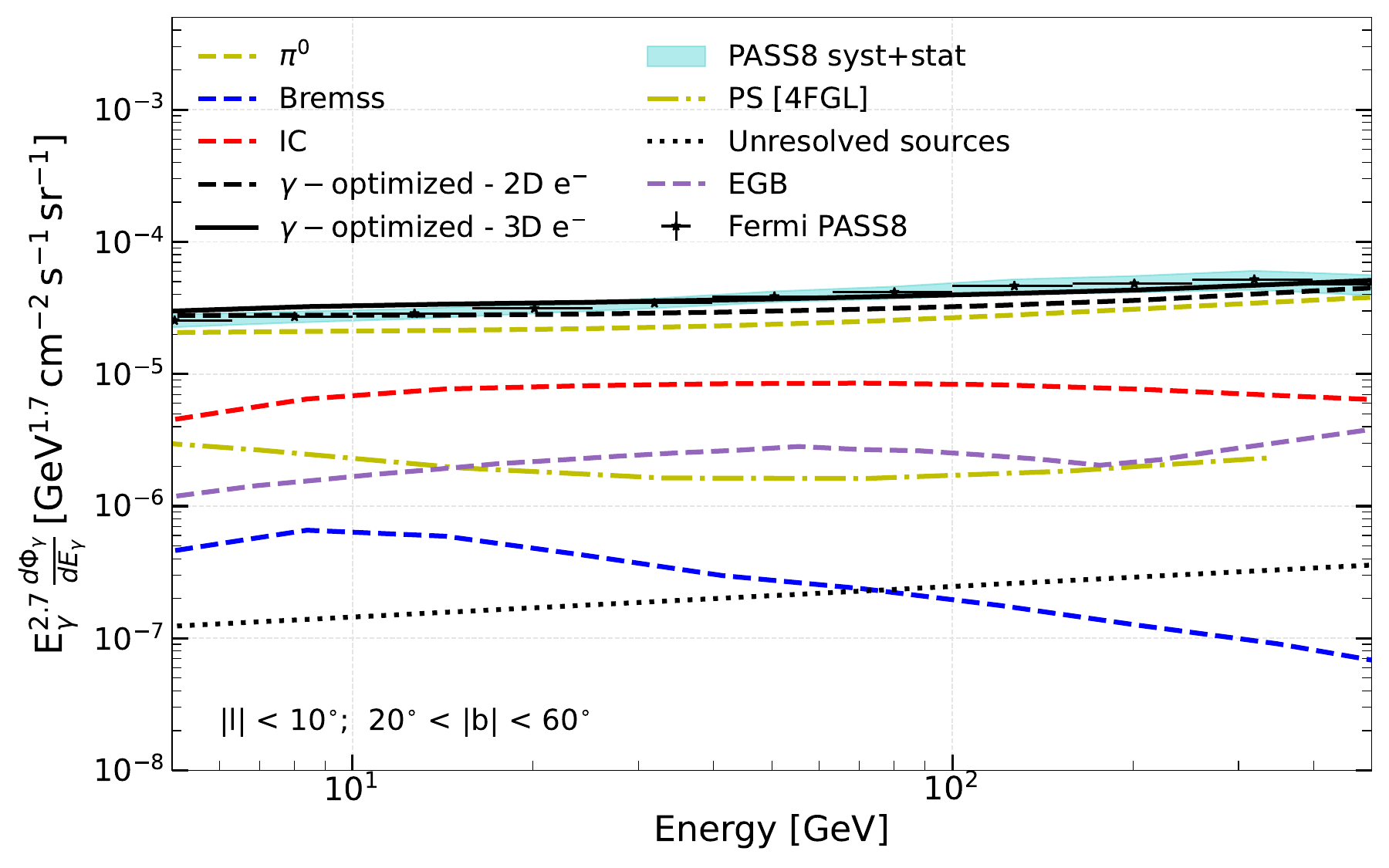} 
\caption{Predicted GDE emission spectra compared with Fermi-LAT data in four different sky windows. Here, we report the different components of the computed emission in different colors, as specified in the legend.}
\label{fig:AbovePlane_App}
\end{figure}

\begin{figure}[ht]
\centering
\includegraphics[width=0.495\textwidth]{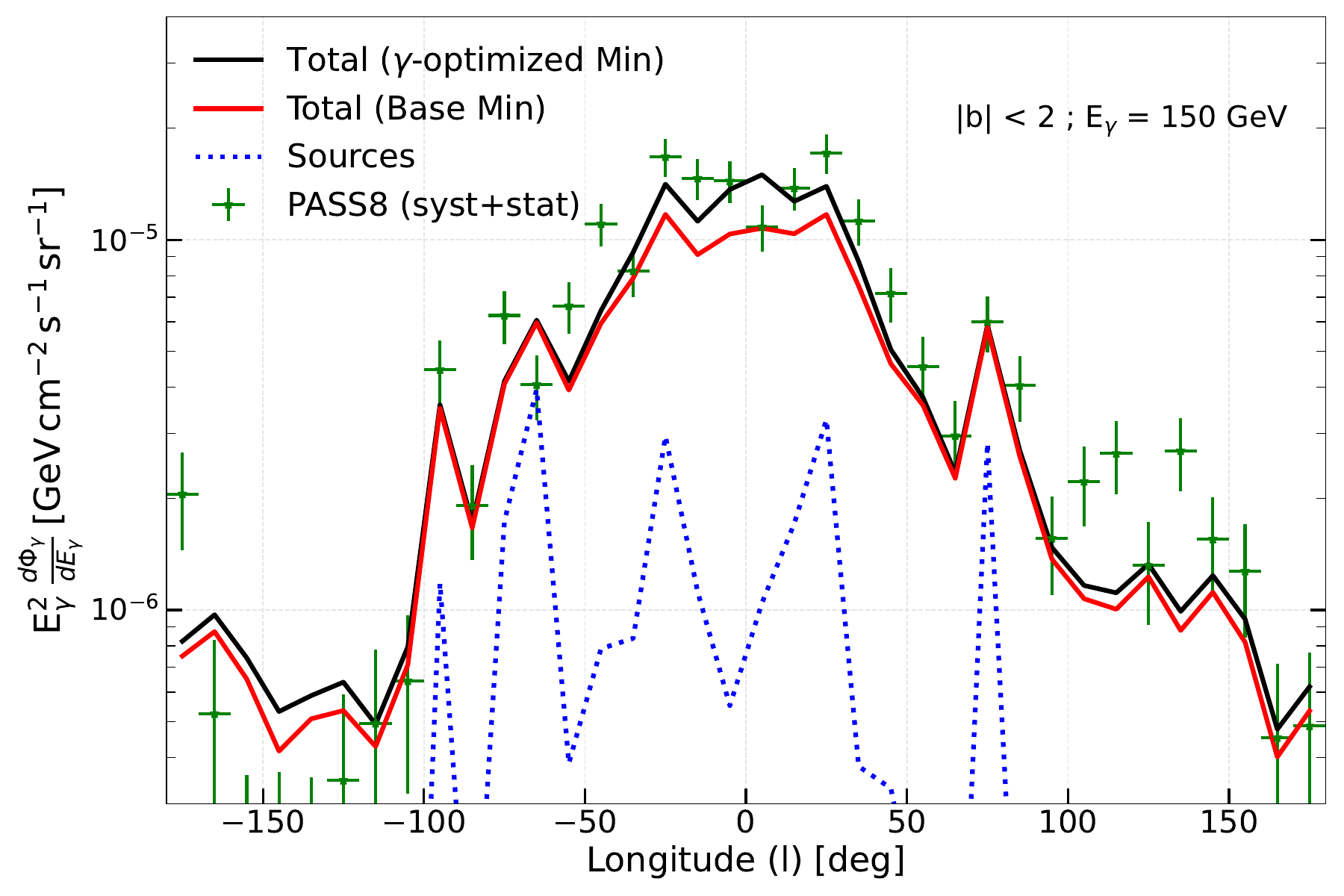}
\includegraphics[width=0.495\textwidth]{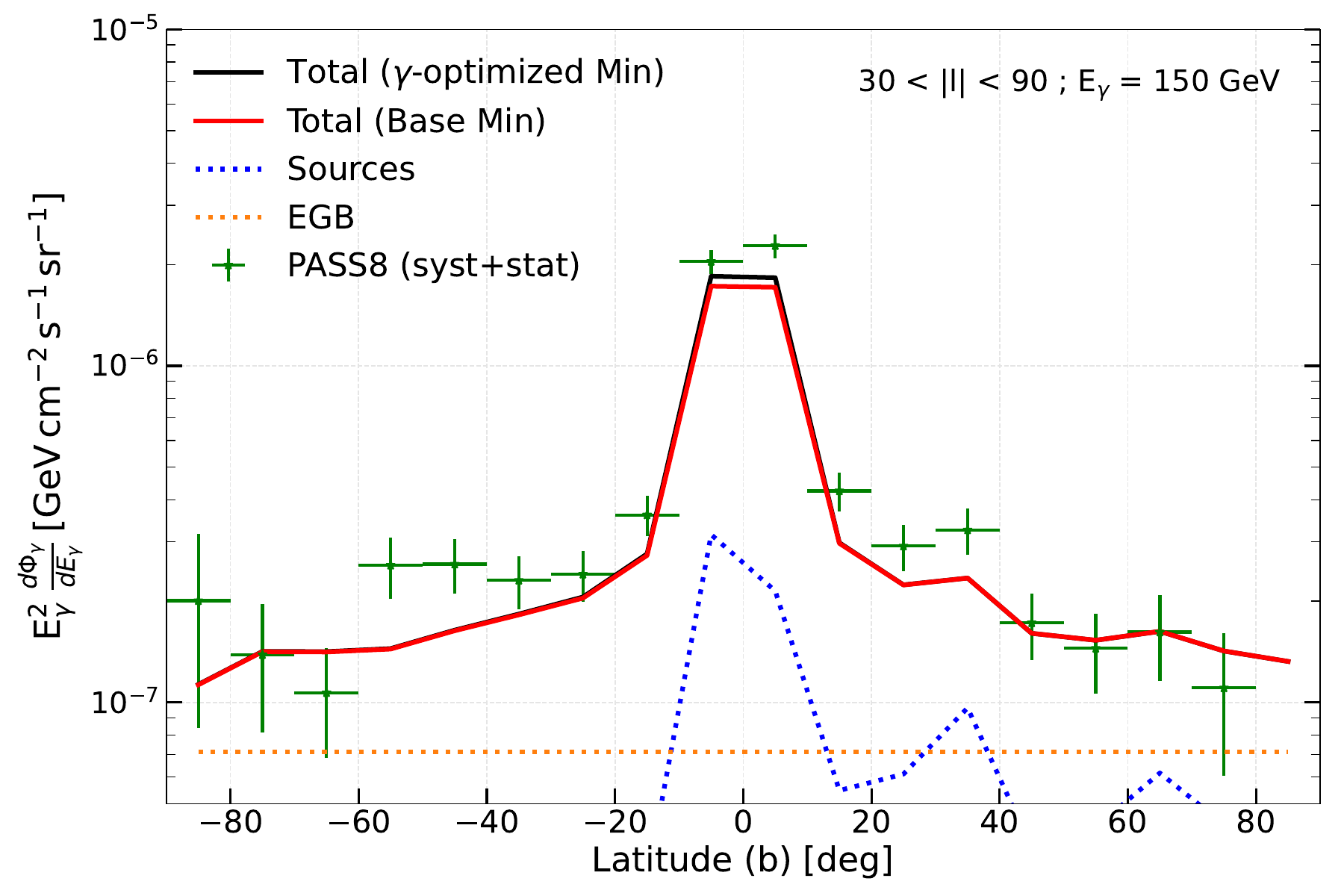}
\caption{Same as in Fig.~\ref{fig:Fermi_profs} but for different sky regions.}
\label{fig:Fermi_profs2}
\end{figure}

Respect to Sec.~\ref{sec:Fermi} here we provide a more detailed comparison between the predictions of our $\gamma$-optimized and Base models, both in the Min setup (at energies below few hundred GeV they roughly coincide with those of the corresponding Max setups), with Fermi-LAT data. 
In Fig.~\ref{fig:AbovePlane_App} we display the different components of the GDE spectrum in the same sky windows as in Fig.~\ref{fig:AbovePlane}. In Fig.~\ref{fig:Fermi_profs2}, we compare the longitude profile and latitude profiles in a different region that the one used in Fig.~\ref{fig:Fermi_profs}.

\newpage
\section{Other comparisons with high-energy gamma-ray data} \label{appendix:HE_comparisons}

\begin{figure}[!h]
\centering
\includegraphics[width=0.49\textwidth]{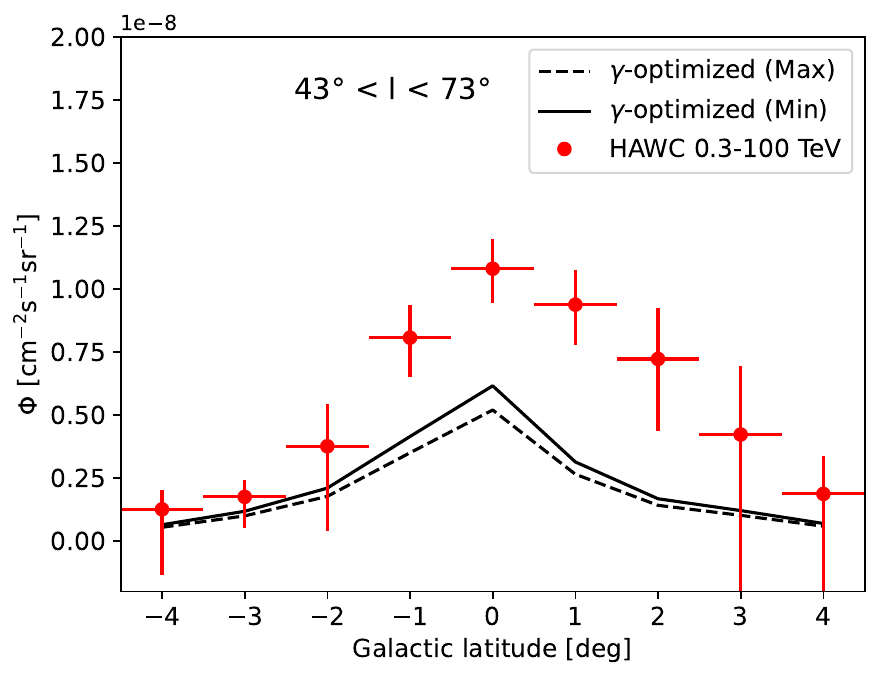}
\includegraphics[width=0.49\textwidth]{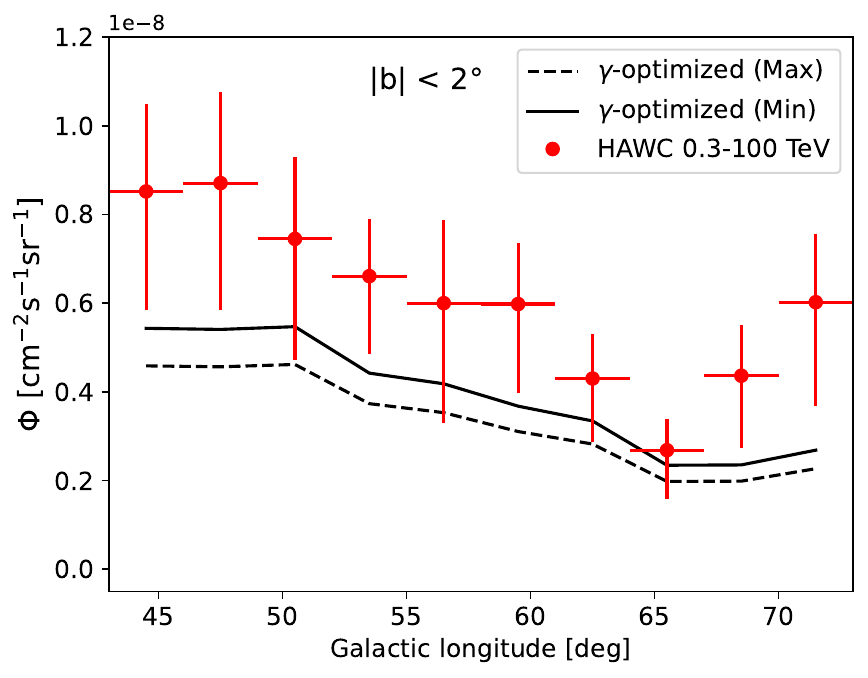}
\caption{The GDE longitude (left) and latitude (right) profiles computed with the $\gamma$-optimized Min and Max models are compared with the HAWC data \cite{HAWC:2023wdq}.
}
\label{fig:Profs_HAWC}
\end{figure}

Extending Secs.~\ref{sec:HAWC_comp}, \ref{sec:Tibet_comp}, \ref{sec:LHAASO_comp}
we report additional comparisons of our $\gamma$-optimized Min and Max models with the GDE longitude and latitude profiles measured by HAWC~\cite{HAWC:2023wdq} between 0.3 and 100 TeV integrated on $- 2^\circ < b < 2^\circ$ and $43^\circ < l < 73^\circ$, respectively, which are shown in Fig.~\ref{fig:Profs_HAWC}.

\begin{figure}[b]
\centering
\includegraphics[width=0.6\textwidth]{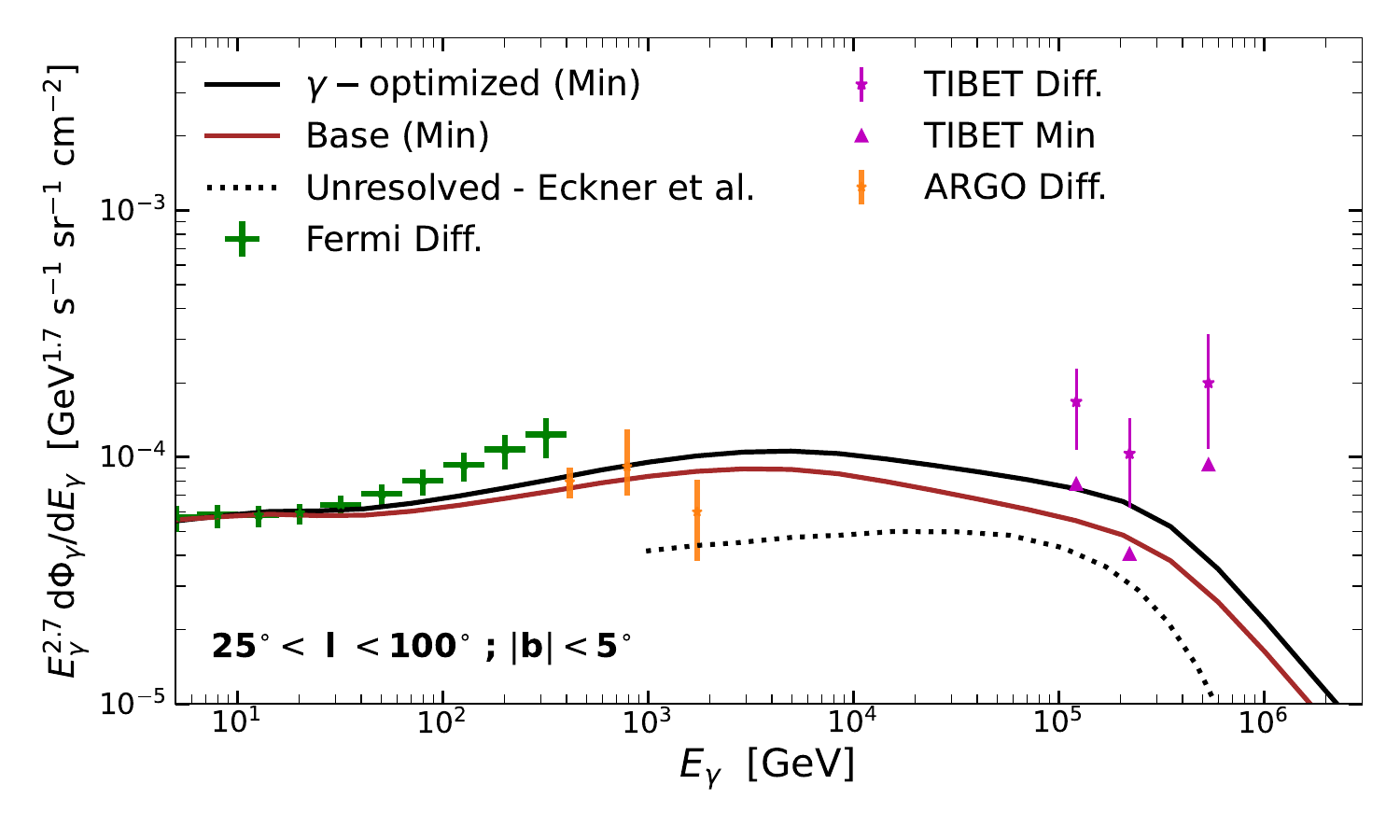}
\caption{Comparison of the Tibet results with the $\gamma$-optimized and Base models both for Min setup. In addition to the truly diffuse emission, represented by the $\gamma$-optimized and Base models, we also show the estimated contribution from unresolved sources, as derived in Ref.~\cite{Eckner_Limits}.}
\label{fig:TIBET_Opt+Base}
\end{figure}

\begin{figure}[t]
\centering
\includegraphics[width=0.495\textwidth]{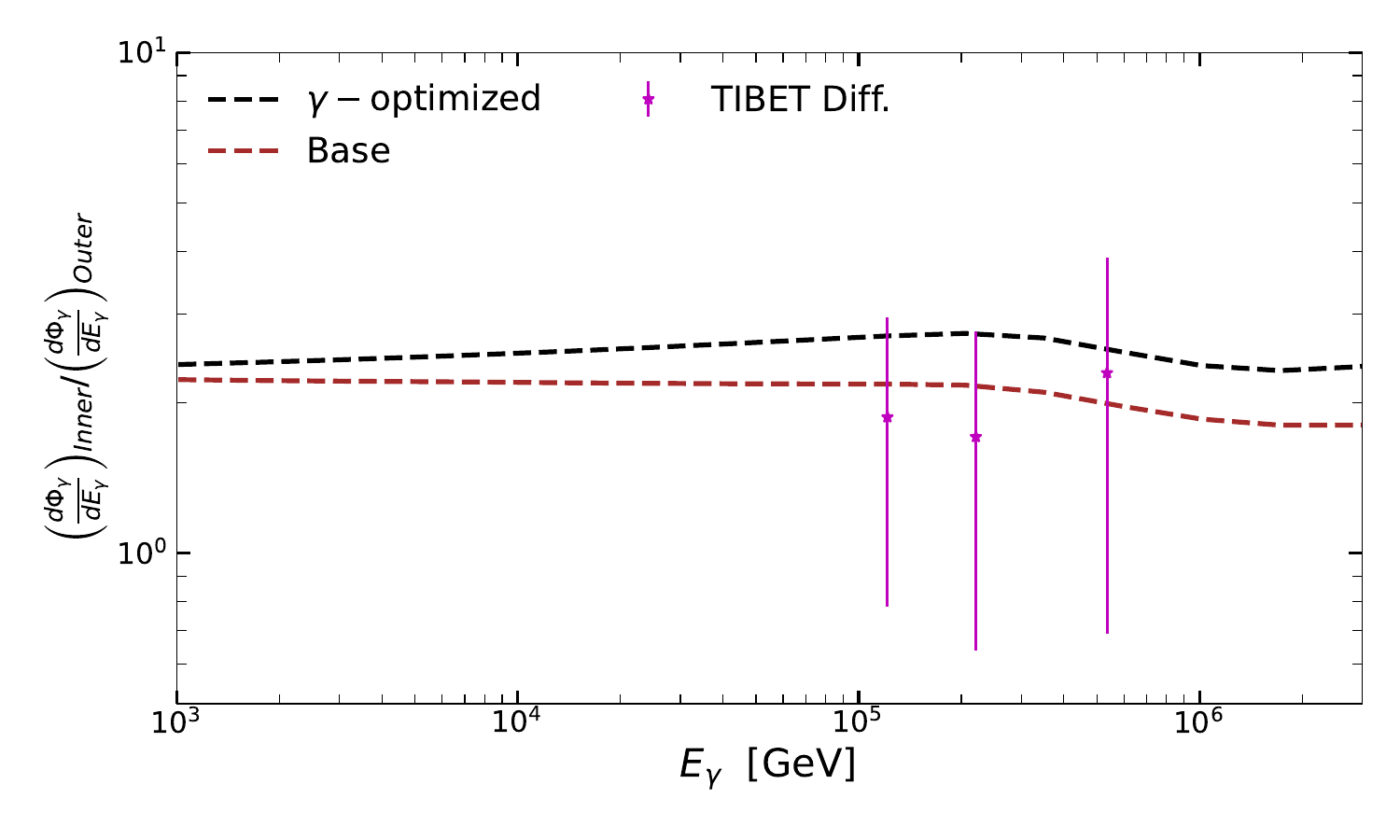}
\includegraphics[width=0.495\textwidth]{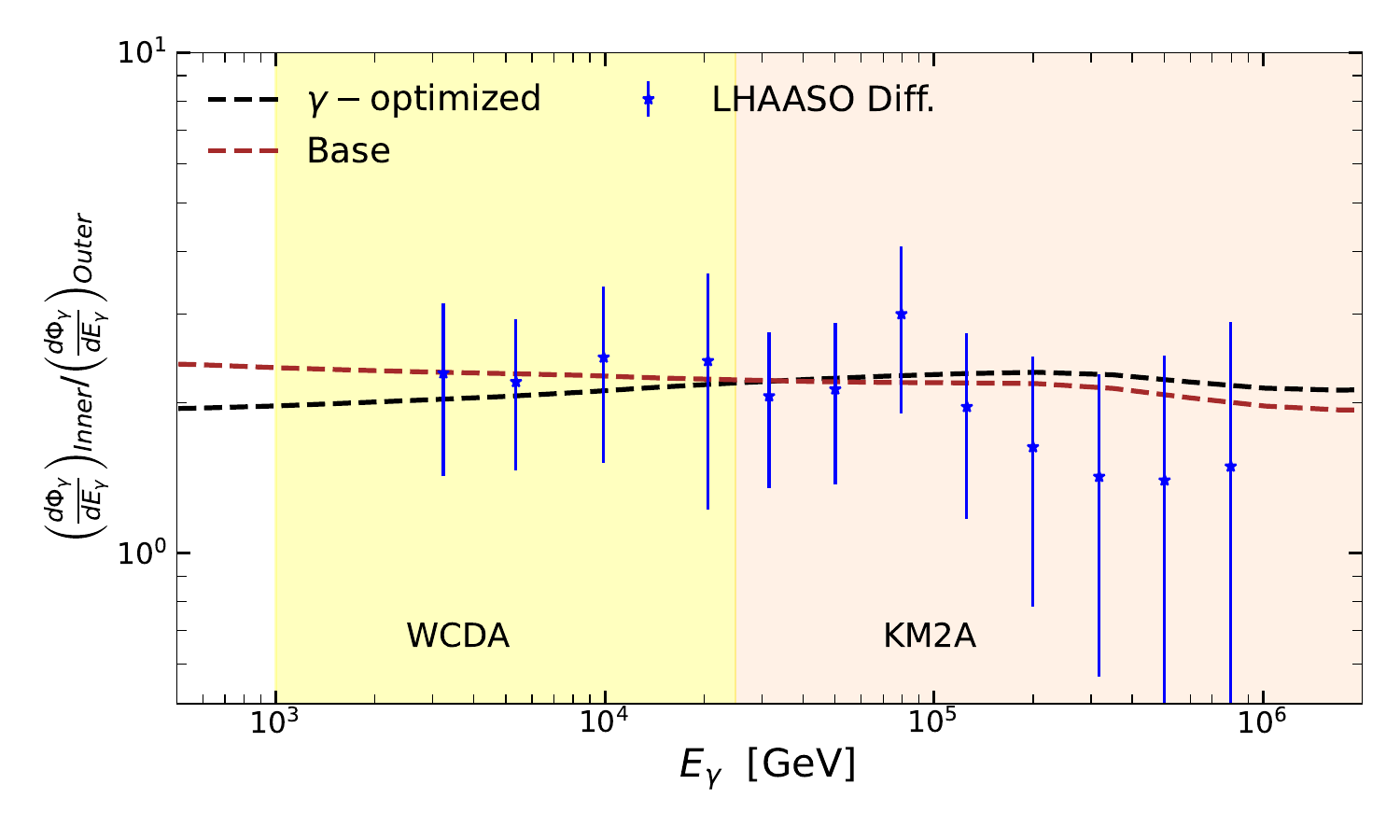}
\caption{Ratio of the observed flux in the inner region over the observed flux in the outer region for Tibet (left) and LHAASO (right). }
\label{fig:Ratios}
\end{figure}

In Fig.~\ref{fig:TIBET_Opt+Base} we compare the GDE spectra computed with the $\gamma$-optimized and Base models, both for the Min setup (favored by LHAASO results), with Tibet-AS$\gamma$ data in Region A. This is different from Fig.~\ref{fig:TIBET_Comp} where $\gamma$-optimized Min and Max setups were rather reported. In this figure for each Tibet data point we also display (triangles) the minimal flux ($1~\sigma$) obtained subtracting the contribution of extended resolved sources as estimated in \cite{Kato:2024ybi}.
The dotted line represents the estimated emission from unresolved sources, as derived by Ref.~\cite{Eckner_Limits}.
No mask is adopted to perform this comparison. 

\begin{figure}[b]
\centering
\includegraphics[width=\textwidth]{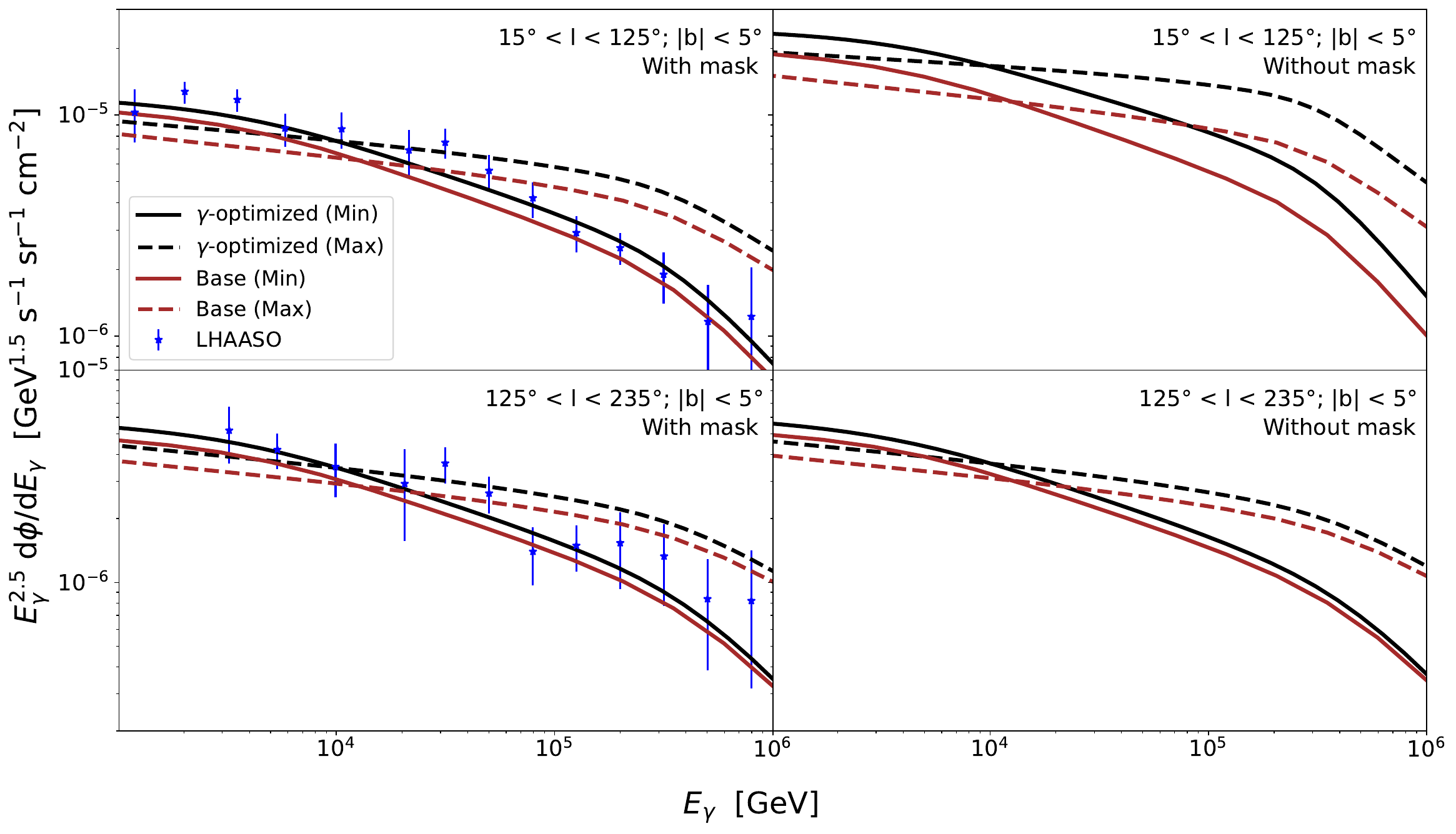}
\caption{We show here the GDE spectra computed with (left panels) and without (right panels) the mask used by LHAASO in the inner and outer regions probed by that experiment.
In the first case the models are compared with LHAASO KM2A and WCDA data.}
\label{fig:Full_LHAASO_Comp}
\end{figure}

Then, in Figure~\ref{fig:Ratios}, we show a comparison of the ratio of the flux in the inner region to the flux in the outer region, as obtained from our $\gamma$-optimized and Base models, compared to that measured by TIBET and LHAASO. Spectral differences between internal and external regions can hint at inhomogeneous diffusion. However, no clear variation is observed from TIBET and LHAASO data. Nevertheless, as we show, even models incorporating inhomogeneous diffusion would not exhibit a very hard trend of this ratio, at least, in comparison to the experimental error bars. In fact, we find that both, uniform and inhomogeneous propagation scenarios seem totally compatible with data. In the case of TIBET, this comparison must be taken with a grain of salt, given that the contribution from unresolved sources may be different in the inner and outer regions and affect this comparison.

Finally, in Fig.~\ref{fig:Full_LHAASO_Comp} we show the GDE predicted for the $\gamma$-optimized and Base models, both for the Min and Max setups, in the inner and outer regions computed with (left panels) and without (right panels) the mask used by the LHAASO collaboration. In the first case a comparison with LHAASO-KM2A and WCDA data is performed.


\section{Additional comparison with available neutrino data}
\label{appendix:neutrino}

\begin{figure}[h]
\includegraphics[width=0.495\textwidth]{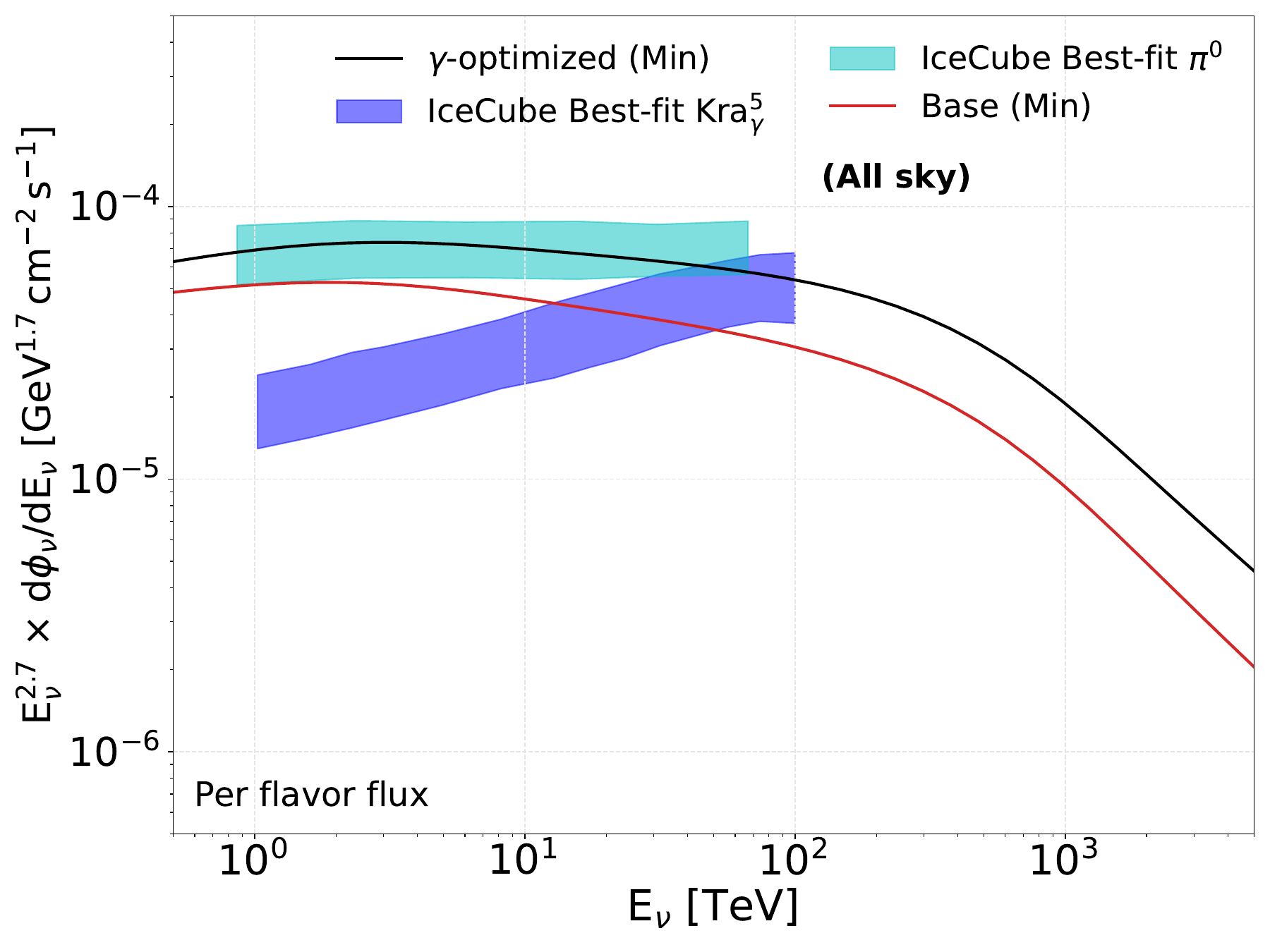} 
\includegraphics[width=0.495\textwidth]{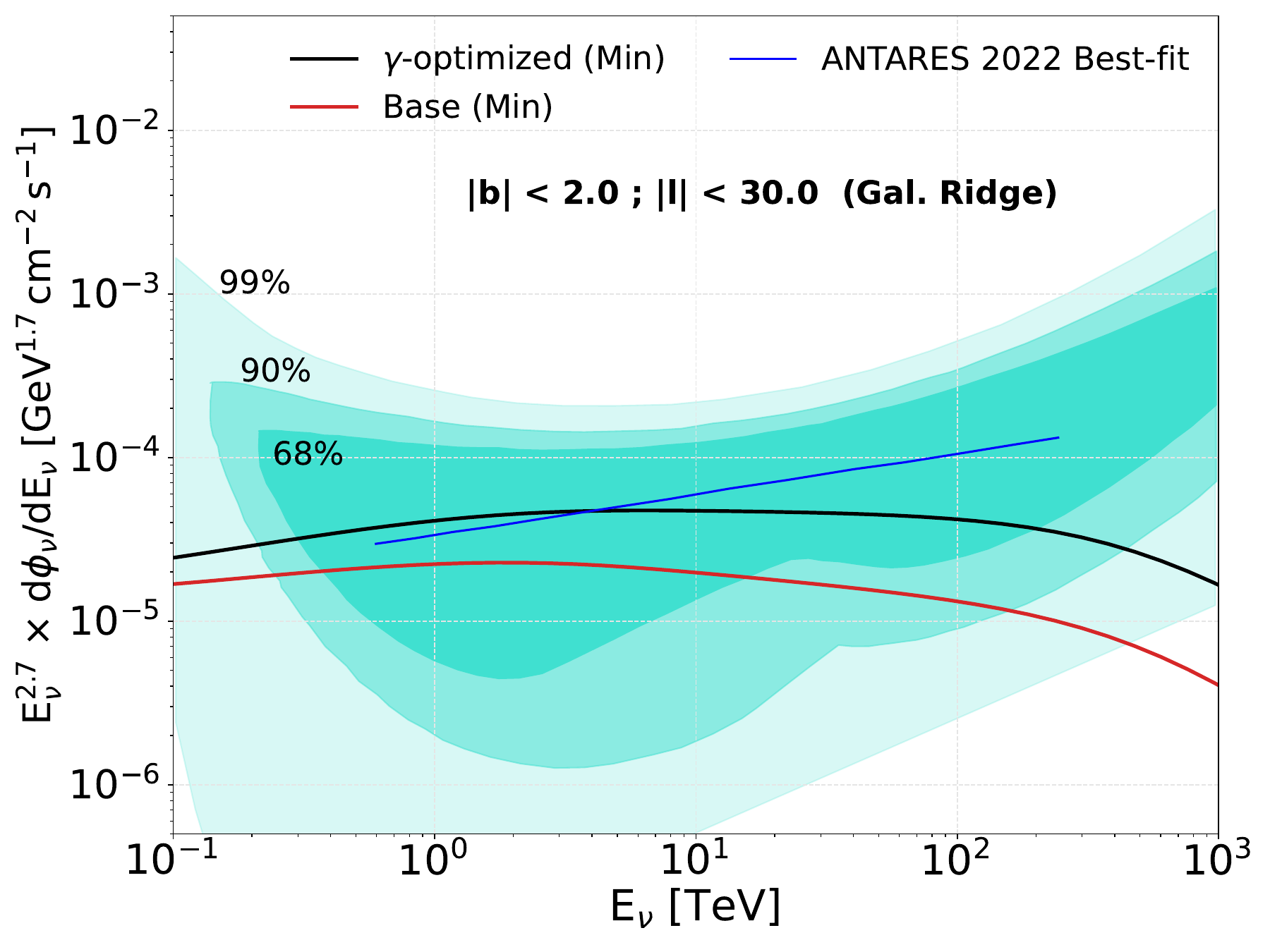}
\caption{Predicted full-sky $\nu$ diffuse emission (per flavor) from the $\gamma$-\textit{optimized} model compared to the Base model. best-fit IceCube flux extracted from the KRA-$\gamma$ (cutoff energy of E$_c=5$~PeV) and $\pi^0$ models. }  
\label{fig:IceCube_App}
\end{figure}

Here, we provide two additional comparisons of our Base and $\gamma$-models (Min) models with neutrino data, similar to Fig.~\ref{fig:IceCube}. In the left panel of Fig.~\ref{fig:IceCube_App} we compare the IceCube results for the $\nu$GDE emission in the whole sky, with the one predicted by our models. We remind the reader that these models only contain the truly diffuse emission (i.e. produced from CR interactions), while a component of Galactic sources is expected to be present at some extent too. We finally present a similar comparison of our models, but with the ANTARES results of the Galactic Ridge, in the right panel of this figure. 

\section{Model parameters}
\label{sec:params}

The injection and propagation of cosmic rays in the models studied here were computed using the {\tt DRAGON2} code \cite{Evoli:2016xgn,Evoli2018jcap} and the computation of the gamma-ray spectrum is performed with the {\tt HERMES} code \cite{Dundovic:2021ryb}. 
These models, presented in Ref.~\cite{Luque:2022buq}, have a halo height of $H=6.7$~kpc \cite{Pedros_1000000th_paper_same_year} and the Sun is located at $R_\odot=8.5$~kpc. The gas content in the Galaxy is described as a series of Galacto-centric rings. The atomic gas is based on $21$~cm line observations by the HI4PI survey \cite{HI_gas_obs}, while the molecular gas is based on CO observations at 115~GHz from the CfA survey \cite{CfA_survey}. A factor $X_{CO}$ is needed in the latter component in order to convert the CO data to H$_2$, and the values adopted are [$1.8$, $3.5$, $4.0$, $4.5$, $7.5$, $8.0$] $10^{20}$~cm$^{-2}$ K$^{-1}$ / (km s$^{-1}$) in the radial intervals [$0 - 3$ kpc; $3 - 5$ kpc; $5 - 6$ kpc; $6 - 7$ kpc; $7 - 15$ kpc; $15 - 30$ kpc]. 

The injection parameters of both H and He are listed in Table~\ref{tab:inj_params}. The Alfv\`en velocity is assumed to be $V_A = 13$~km s$^{-1}$. The diffusion coefficient is described as
\begin{equation}
    D(\rho,\mathbf{x}) = D_0 \cdot \left(\frac{\rho}{\rho_0}\right)^{\delta(R)},
\end{equation}
where $D_0 = 6.1 \cdot 10^{28}$~cm$^2$s$^{-1}$, $\rho_{0}= 4$~GV, $\beta$ is the velocity of the particles and $R$ the Galactocentric radius. 
The spatial-dependent diffusion index $\delta(R)$ is set to match the hardening of Fermi-LAT $\gamma$-ray data towards the GC \cite{Gaggero:2014xla,Gaggero:2015xza,Fermi-LAT:2016zaq,Yang2016prd,Lipari:2018gzn,Pothast:2018bvh} and takes the form 
\begin{equation}
    \delta(R) = 0.04({\rm kpc^{-1}}) \cdot R({\rm kpc}) + 0.17,
\end{equation}
for $R < R_\odot$ and $\delta(R) = \delta(R_\odot)$ for $R \geq R_\odot$.
In the Base models, the diffusion coefficient is characterized by a  value of $\delta = 0.5$ constant in the whole Galaxy, with the same normalization quoted above.

\begin{table*}[t]
    \centering
    \begin{tabular}{|c|c|c|c|c|c|c|c|c|}
    \hline
    \multicolumn{9}{c}{\textbf{Injection parameters}} \\
    \hline 
    \hline
    & $^1\mathbf{H} \; \gamma_1$ & $^1\mathbf{H} \; \gamma_2$ & $^1\mathbf{H} \; \gamma_3$ & $^1\mathbf{H} \; \gamma_4$ & $^4\mathbf{He} \; \gamma_1$ & $^4\mathbf{He} \; \gamma_2$ & $^4\mathbf{He} \; \gamma_3$ & $^4\mathbf{He} \; \gamma_4$ \\
    \hline
    Max model & 2.33 & 2.23 & 2.78 & --- & 3.28 & 2.18 & 2.69 & --- \\
    \hline
    Min model & 2.33 & 2.16 & 2.44 & 3.37 & 2.30 & 2.06 & 2.34 & 3.01 \\
    \hline
  \end{tabular}
  
  \caption{\small{Spectral indexes at injection for the Max and Min models. These spectral indexes are tuned to CR local data as described above and correspond to spectral breaks at the following energies: $335$ and $6 \cdot 10^6$~GeV for the Max models and $335$, $2\cdot 10^4$ and $4 \cdot 10^6$~GeV for the Min models. }}
    \label{tab:inj_params}
\end{table*}

\bibliographystyle{apsrev4-1}
\bibliography{biblio}

\end{document}